\documentclass[12pt]{article}

\usepackage{rotate}
\usepackage{epsf}
\usepackage{psfig}

\begin{document} 

\epsfclipon

\title{\bf FLAVOR ASYMMETRY OF THE SEA QUARK DISTRIBUTIONS}
\author{{\it Jen-Chieh Peng and Gerald T. Garvey}}
\maketitle

\begin{center}
Physics Division\\
Los Alamos National Laboratory\\
Los Alamos, NM 87545 USA
\end{center}

\bigskip

\noindent{CONTENTS}

\noindent{1. INTRODUCTION}

\noindent{2. EARLY STUDIES OF THE NUCLEON SEA}

2.1 Deep Inelastic Scattering

2.2 Gottfried Sum Rule

2.3 Drell-Yan Process

\noindent{3. RECENT EXPERIMENTAL DEVELOPMENTS}

3.1 NMC Measurements of the Gottfried Integral

3.2 E772 Drell-Yan Experiment

3.3 NA51 Drell-Yan Experiment

3.4 E866 Drell-Yan Experiment

3.5 HERMES Semi-Inclusive Experiment

3.6 Impact on the Parton Distribution Functions

3.7 Comparison Between Various Measurements

\noindent{4. ORIGINS OF THE $\bar d/ \bar u$ ASYMMETRY}

4.1 Pauli Blocking

4.2 Meson-Cloud Models

4.3 Chiral Models

4.4 Instanton Models  

4.5 Charge Symmetry Violation

\noindent{5. FURTHER IMPLICATIONS OF THE MESON-CLOUD MODELS}

5.1 Strange Sea of the Nucleon

5.2 Spin Dependent Structure Functions

5.3 Sea Quark Distributions in Hyperons

\noindent{6. FUTURE PROSPECTS}

6.1 $\bar d / \bar u$ at Large and Small $x$

6.2 W Production

6.3 Strange Sea in the Nucleon

6.4 Sea Quark Polarization

\noindent{7. CONCLUSION}

\vskip 0.5in

\begin{abstract}
Recent deep inelastic scattering and Drell-Yan experiments have revealed 
a surprisingly large asymmetry between the up and down sea quark
distributions in the nucleon. This result strongly suggests that the mesonic
degrees of freedom play an important role in the description of the parton
distributions in hadrons. In this article, we review the current status of 
the flavor asymmetry of the nucleon sea. The implications of
the pionic models as well as possible future measurements are also discussed.
\end{abstract}

\newpage
\noindent{\bf{1. INTRODUCTION}}

The first direct evidence for point-like constituents in the nucleons came 
from the observation of scaling phenomenon in Deep-Inelastic Scattering
(DIS) experiments~\cite{bloom69,breiden69} at SLAC. 
These point-like charged constituents, called 
partons, were found to be spin-1/2 fermions. These partons were initially
identified as the quarks in the Constituent
Quark Models (CQM). However, it was soon realized that valence 
quarks alone could not account for the large enhancement of the cross
sections at small Bjorken-$x$, the fraction of nucleon momentum carried
by the partons. These swarm of low-momentum partons,
dubbed wee-partons by Feynman~\cite{feyn72}, was 
interpreted as the quark and antiquark
sea of the nucleon. DIS experiments therefore provided the first evidence
for the existence of antiquarks in the nucleon.

The observation of partons in DIS experiments paved the road to the 
formulation of Quantum Chromodynamics (QCD) as the theory for strong 
interaction. Nevertheless, the exact form of the parton distribution 
functions can not be deduced from perturbative QCD. Like many other static 
properties of hadrons, the parton distribution functions belong to the 
domain of non-perturbative QCD. In spite of great progress~\cite{liu99} made 
in Lattice Gauge Theory (LGT) in treating the bound-state properties of 
hadrons, it remains a challenge to predict the parton distributions 
using LGT. 

Until parton distributions can be readily calculated from first principles, 
they are best determined from experiments. Electroweak processes such
as DIS and lepton-pair production provide the cleanest means to extract 
information on the parton distributions. There are at least two reasons
why it is important to measure the parton distribution functions. First,
the description of hard processes in high energy interactions requires parton 
distribution functions as an essential input. Second, many aspects 
of the parton distributions, such as sum rules, scaling-violation,
asymptotic behaviors at large and small $x$, and flavor and spin 
structures, can be compared with the predictions of perturbative as well
as non-perturbative QCD.

In this article, we review the status of our current knowledge on the
flavor dependence of the sea quark distributions in hadrons. 
The recent observation of a striking flavor asymmetry
of the nucleon sea has profound implications
on the importance of meson degrees of freedom for the description of parton
substructures in the nucleon. In Section 2, we review the early studies
of the nucleon sea in DIS and lepton-pair production. The crucial recent 
experiments establishing the up/down flavor asymmetry of the nucleon sea
are discussed in Section 3. The various theoretical models for 
explaining the $\bar u/ \bar d$ asymmetry are described in Section 4.
The implications of meson-cloud models on other aspects of the
parton structure functions are discussed in Section 5. Finally, we
present future prospects in Section 6, followed by 
conclusion in Section 7.\\

\noindent{\bf{2. EARLY STUDIES OF THE NUCLEON SEA}}

\noindent{\bf{2.1 Deep Inelastic Scattering}}

Although scaling behavior in inelastic electron scattering
was predicted by Bjorken~\cite{bj69} based on the framework of
current algebra, its confirmation by the SLAC experiments still came as a
major surprise. A simple and intuitive picture for explaining the
scaling behavior is the parton model advanced by 
Feynman~\cite{feyn72,feyn69}. In this model,
the electron-nucleon deep-inelastic scattering is described 
as an incoherent sum of elastic electron-parton scattering. However,
the nature of the partons within the nucleon was not specified by
Feynman. Some authors~\cite{drell69,drell70,cab70,lee72} speculated 
that the partons are the `bare nucleon'
plus the pion cloud, while others~\cite{bj69a,kuti71,land71} believed 
they are the quarks
introduced by Gell-Mann~\cite{gell64} and Zweig~\cite{zweig64}. 
The latter scenario was strongly
supported by the measurement of $R$, ratio of the longitudinally
over transversely polarized photon cross sections, showing the spin-1/2
character of the partons.

Evidence for quark-antiquark sea 
in the nucleon came from the observation that the structure function
$F_2(x)$ approaches a constant value as $x \to 0$~\cite{friedman}. 
If the proton is made 
up of only three quarks, or any finite number of quarks, $F_2(x)$ is expected
to vanish as $x \to 0$. Bjorken and Paschos~\cite{bj69a} 
therefore assumed that the nucleon consists of three quarks in a 
background of an indefinite number of quark-antiquark pairs.
Kuti and Weisskopf~\cite{kuti71} further 
included gluons among the constituents of nucleons
in order to account for the missing momentum not carried by the
quarks and antiquarks alone.

The importance of the quark-antiquark pairs in the nucleon is in sharp 
contrast to the situation for the atomic system, where 
particle-antiparticle pairs play a relatively minor role (such as the 
polarization of the vacuum). In strong interactions, pairs of virtual
fermions or virtual Goldstone bosons are readily produced as a result of the
relatively large magnitude of the coupling constant $\alpha_s$,
and they form an integral part of the nucleon's structure.

Neutrino-induced DIS experiments allowed the separation of sea quarks
from the valence quarks. Recall that
\begin{eqnarray}
F_2^{\nu p}(x) = 2x \sum_i \left[q_i(x) + \bar q_i(x)\right], \hspace{0.8in} 
\nonumber \\ F_3^{\nu p}(x) = 2 \sum_i \left[q_i(x) - \bar q_i(x)\right] 
= 2 \sum_i q_i^v(x),
\label{eq:2.1.1}
\end{eqnarray}
where $i$ denotes the flavor of the quarks.
Note that the valence quark distribution is defined as the difference
of the quark and antiquark distributions, $q^v_i(x) = q_i(x) - \bar q_i(x)$.
Eq.~\ref{eq:2.1.1} shows that the valence quark distribution is simply 
$F_3^{\nu p}(x)/2$, while the sea quark distribution is given by
$F_2^{\nu p}(x)/2x - F_3^{\nu p}(x)/2$. 
The $F_2(x)$ and $F_3(x)$ data from the CDHS experiment~\cite{cdhs79}
clearly showed that the valence 
quark distributions dominate at $x>0.2$, while the sea quarks are 
at small $x$. 

The earliest parton models assumed that the proton sea was flavor symmetric,
even though the valence quark distributions are clearly flavor asymmetric.
Inherent in this assumption is that the content of the sea is 
independent of the valence quark's composition. Therefore,
the proton and neutron are expected to have identical sea-quark distributions.
The flavor symmetry assumption was not based on any known physics, and 
it remained to be tested by experiments. Neutrino-induced charm production
experiments~\cite{abrom82,conrad98}, which are 
sensitive to the $s \to c$ process, provided strong
evidences that the strange-quark content of the nucleon is only about half
of the up or down sea quarks. Such flavor asymmetry is attributed to the
much heavier strange-quark mass compared to the up and down quarks. The similar
masses for the up and down quarks suggest that the nucleon sea should be nearly
up-down symmetric. A very interesting method
to check this assumption is to measure the Gottfried integral 
in DIS, as discussed next.\\

\noindent{\bf{2.2 Gottfried Sum Rule}}

In 1967, Gottfried studied electron-proton scattering assuming that 
the proton consists of three constituent quarks~\cite{gott67}. He 
showed that the total
electron-proton cross section (elastic plus inelastic) is identical to 
the Rutherford scattering from a point charge. Gottfried derived a sum rule
\begin{eqnarray}
I^p_2 = \int_0^1 F^p_2 (x,Q^2)/x~ dx = \sum_i (Q^{p}_{i})^2 = 1,
\label{eq:2.2.1}
\end{eqnarray}
where $Q^{p}_{i}$ is the charge of the $i$th quark in the proton.
Gottfried expressed great skepticism that this sum rule would be confirmed by
the forthcoming SLAC experiment by stating ``I think Prof. Bjorken and
I constructed the sum rules in the hope of destroying 
the quark model''~\cite{bj67}.
Indeed, Eq.~\ref{eq:2.2.1} was not confirmed by the 
experiments, not because of the
failure of the quark model, but because of the presence of quark-antiquark sea.
In fact, the infinite number of the sea partons makes $I^p_2$ 
diverge. A closely related sum rule, now called the Gottfried Sum Rule (GSR),
avoids this problem by considering the difference
of the proton and neutron cross sections, namely,
\begin{eqnarray}
I^p_2 - I^n_2 = \int_0^1 [F^p_2 (x,Q^2)-F^n_2 (x,Q^2)]/x~ dx 
= \sum_i [(Q^{p}_{i})^2 - (Q^{n}_{i})^2] = 1/3.
\label{eq:2.2.2}
\end{eqnarray}
In deriving Eq.~\ref{eq:2.2.2}, it was assumed 
that the sea quarks in the proton and 
neutron are related by charge symmetry (see Section 4.5).

Soon after the discovery of scaling in electron-proton DIS,
electron-deuterium scattering experiments were carried out to
extract the electron-neutron cross sections. The comparison of
$e-p$ with $e-n$ data was very important for distinguishing early competing
theoretical models~\cite{friedman}. These data also allowed a
first evaluation~\cite{bloom70} of the Gottfried integral
in 1970. The
first result for the Gottfried integral was 0.19, considerably less than 1/3.
Note that the data only covered $x > 0.08$, and it was assumed that 
$F^p_2 - F^n_2$ follows Regge behavior (proportional to $x^{1/2}$)
in the unmeasured small-$x$ region. Due to the $1/x$ factor in the integrand, 
the small-$x$ region could have potentially large contributions to
the Gottfried integral. Moreover, it was not clear if $F^p_2 - F^n_2$ would
indeed follow the Regge behavior at small $x$, and if so, at what value of
$x$. By 1973, new data were available down to $x = 0.05$ and the Gottfried
integral was evaluated to be 0.28~\cite{bloom73}, considerably larger
than the first result. 
It should be pointed out that these data were taken at relatively low
values of $Q^2$. Furthermore, $Q^2$ varies as a function
of $x$. 

Although the large systematic errors associated with 
the unmeasured small-$x$ region prevented
a sensitive test of the GSR, Field and 
Feynman~\cite{field77} nevertheless
interpreted the early SLAC data as strong indications that GSR is
violated and that the $\bar u$ and $\bar d$ distributions in the proton 
are different.
The relationship between the Gottfried integral and the 
$\bar d/ \bar u$ asymmetry is clearly seen in the parton model, namely,
\begin{eqnarray}
\int_0^1 [F^p_2 (x,Q^2)-F^n_2 (x,Q^2)] /x~ dx 
= {1 \over 3} + {2 \over 3}
\int_0^1 [\bar u(x) - \bar d(x)] dx.
\label{eq:2.2.2.1}
\end{eqnarray}
Eq.~\ref{eq:2.2.2.1} clearly shows that the early SLAC data on GSR
implied $\bar d > \bar u$, at least for certain region of $x$.
Field and Feynman further suggested that Pauli blocking
from the valence quarks would inhibit the $\bar u u$ sea more than the
$\bar d d$ sea, hence creating an asymmetric nucleon sea.

\begin{figure}
\center
\psfig{figure=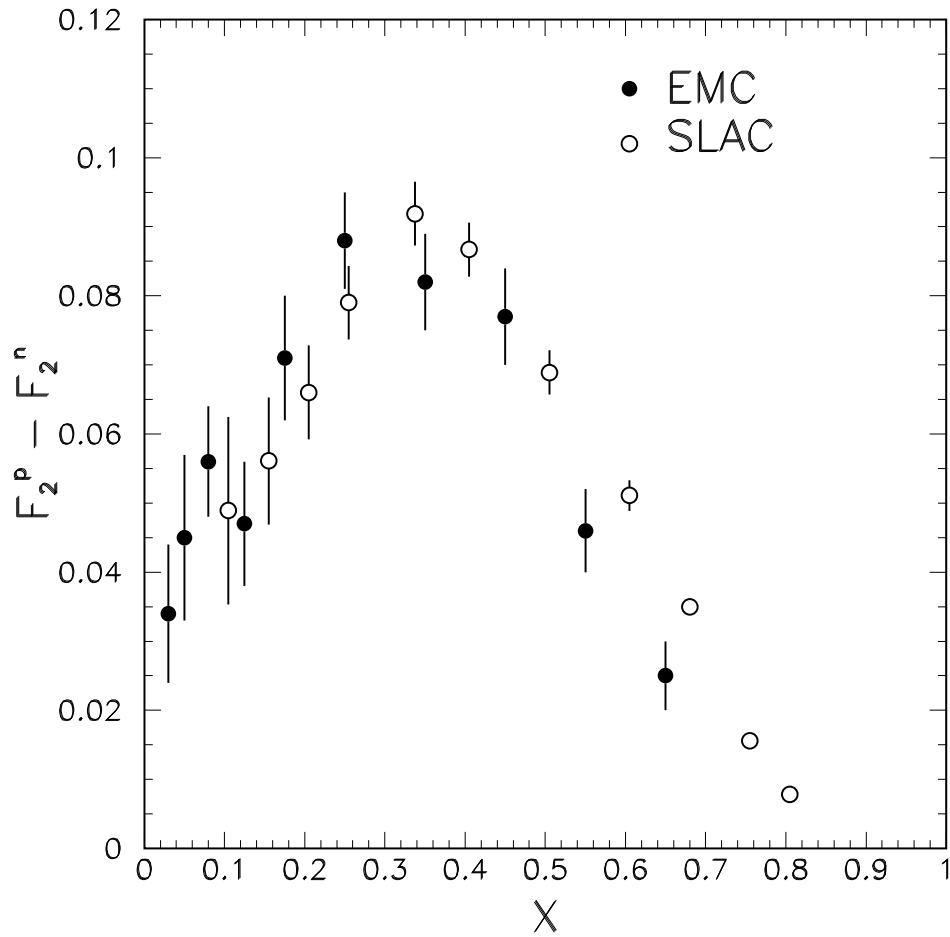,height=5.5in}
\caption{EMC~\cite{emc87} and SLAC~\cite{bodek79} measurements 
of $F^p_2 - F^n_2$.}
\label{fig:2.2.1}
\end{figure}

The SLAC DIS experiments were followed by several muon-induced DIS experiments
at Fermilab and at CERN. Using high energy muon beams, these
experiments reached much larger values of $Q^2$ and they
clearly established~\cite{wate75,chang75,chio79} the scaling-violation 
phenomenon in DIS. 
The Gottfried integral was also evaluated in muon DIS 
experiments~\cite{emc87,bcdms90}. Figure~\ref{fig:2.2.1}
compares the data from the European Muon 
Collaboration (EMC)~\cite{emc87} with earlier
electron data from SLAC~\cite{bodek79}. The coverages 
in $x$ are similar in these two
experiments, even though
the $Q^2$ values covered by the EMC are much larger.
Figure~\ref{fig:2.2.1} shows that $F^p_2 -F^n_2$ from EMC tend to shift towards 
smaller $x$ relative to the SLAC data, in qualitative agreement 
with the QCD $Q^2$-evolution. 
The Gottfried integral 
determined from the EMC experiment is $0.235 + 0.110 - 0.099$, consistent
with that from the SLAC, but still lower than 1/3. 

Despite the fact that all measurements of Gottfried integral consistently showed
a value lower than 1/3, the large systematic errors prevented a
definitive conclusion. As a result, all 
parametrizations~\cite{do84,ehlq84,dflm88,mrs88,abfow89}
of the parton distributions based on global fits to existing data before 1990 
assumed a symmetric $\bar u$, $\bar d$
sea. As discussed later, compelling evidence for an asymmetric light-quark
sea awaited new experimental inputs.\\

\noindent{\bf{2.3 Drell-Yan Process}}

The first high-mass dilepton production experiment~\cite{chris70} 
was carried out at the 
AGS in 1969, soon after scaling was discovered at SLAC.
Drell and Yan~\cite{dy71} interpreted the 
data within the parton model, in which 
a quark-antiquark pair annihilate into a virtual
photon decaying subsequently into a lepton pair. This simple model was
capable of explaining several pertinent features of the data, including the
overall magnitude of the cross sections, the scaling behavior of the cross
sections, and the polarization of the virtual photons. The high-mass continuum
lepton-pair production is therefore called the Drell-Yan (DY) process.

Since the underlying mechanism for the DY process involves the annihilation
of a quark with an antiquark, it is not surprising that this process can
be used to probe the antiquark contents of the beam or target hadrons.
In the parton model, the DY cross section is given by
\begin{eqnarray}
{d^2\sigma\over dM^2dx_F}={4\pi\alpha^2\over 9M^2s}{1\over (x_1+x_2)}\sum_a
e_a^2[q_a(x_1)
\bar q_a(x_2)+\bar q_a(x_1)q_a(x_2)]. 
\label{eq:2.3.1}
\end{eqnarray}
Here $q_a(x)$ are the quark or antiquark structure functions of 
the two colliding hadrons evaluated at momentum fractions $x_1$ and 
$x_2$. The sum is over quark flavors.  
In addition, one has the kinematic relations,
\begin{eqnarray}
& &\tau\equiv x_1x_2 = M^2/s,\nonumber \\
& &x_F = x_1-x_2, 
\label{eq:2.3.2}
\end{eqnarray}
where $M$ is the invariant mass of the lepton pair and $s$ is the square 
of the center-of-mass energy. The cross section is 
proportional to $\alpha^2$, indicating its electromagnetic character. 

Eq.~\ref{eq:2.3.1} shows that the 
antiquark distribution enters as a multiplicative
term in the DY cross section rather than an additive term in the
DIS cross section. Hence, the antiquark
distributions can be sensitively determined in the DY experiments. The 
dimuon data from the FNAL E288, in which the $\Upsilon$ resonances were 
discovered~\cite{herb77}, were analysed~\cite{kaplan77} to 
extract the sea quark distribution
in the nucleon. By assuming a flavor-symmetric nucleon sea, namely,
$\bar u(x) = \bar d(x) = \bar s(x) = Sea(x)$, the dimuon
mass distribution obtained in 400 GeV proton-nucleus interaction was  
described by $xSea(x) = 0.6 x^{-10}$.
In a later analysis~\cite{ito81} containing 
additional data at 200 and 300 GeV,
E288 collaboration found that a much better fit could be obtained with an
asymmetric sea, namely,
\begin{eqnarray}
&\bar u(x) = (1-x)^{3.48} \bar d(x),~ ~ ~ ~\bar s(x) = (\bar u(x) + \bar d(x))/4.
\label{eq:2.3.3}
\end{eqnarray}
The need for an asymmetric $\bar u$ and $\bar d$ was also manifested
in the E288 $d\sigma/dy$ data~\cite{ito81} at $y = 0$, where $y$ is 
the center-of-mass rapidity. For $p + A$ collision, $d\sigma/dy$ 
at $y = 0$ is expected to
be positive due to the excess of $u$ over $d$ valence quarks in the proton.
The E288 data showed that the slopes are indeed positive, 
but larger than expected from a flavor symmetric sea.
A surplus of $\bar d$ 
over $\bar u$ in the proton sea
would lead to more positive slope in agreement with the data~\cite{ito81}.

The FNAL E439 collaboration~\cite{smith81} studied high 
mass dimuons produced in $p + W$ interaction at 400 GeV.
Their spectrometer covered a considerably larger range
in $x_F$ than E288. They again found that an asymmetric sea, 
$\bar u(x) = (1-x)^{2.5} \bar d(x) $, can well describe their data.

With all the tantalizing evidence for an asymmetric sea
from DIS and DY experiments, it is curious
that all global analyses~\cite{do84,ehlq84,dflm88,mrs88,abfow89} 
of parton distributions in the 1980's still
assumed a symmetric light-quark sea. This probably reflected the 
reluctance to adopt an unconventional description of the nucleon sea without
compelling and unambiguous experimental evidences. As discussed in the
next Section, such evidence became available in the 1990's.\\

\noindent{\bf{3. RECENT EXPERIMENTAL DEVELOPMENTS}}

\noindent{\bf{3.1 NMC Measurements of the Gottfried Integral}}

After the discovery of the so-called `EMC effect'~\cite{emc83} 
which showed that the parton
distribution in a nucleus is different from that in the deuteron, the EMC
detector system was modified by the New Muon Collaboration (NMC) to
study in detail the EMC effect. Special emphases were placed by the NMC on the
capability to explore the 
low-$x$ region where the `shadowing effect'~\cite{emc88} 
is important,
and on accurate measurement of cross section ratios~\cite{nmc90}.
By implementing a `small angle' trigger which extended the 
scattering angle coverage down to 5 mrad, the lowest value of $x$ reached 
by NMC was $\sim 0.001$. The NMC also placed two targets in the
muon beam, allowing DIS data from two different targets to be recorded
simultaneously, thus greatly reducing the beam flux normalization uncertainty.
To account for the different geometric acceptances for events originating from
the two targets, half of the data were taken using a different target 
configuration where the locations of the two targets were interchanged.

\begin{figure}
\center
\psfig{figure=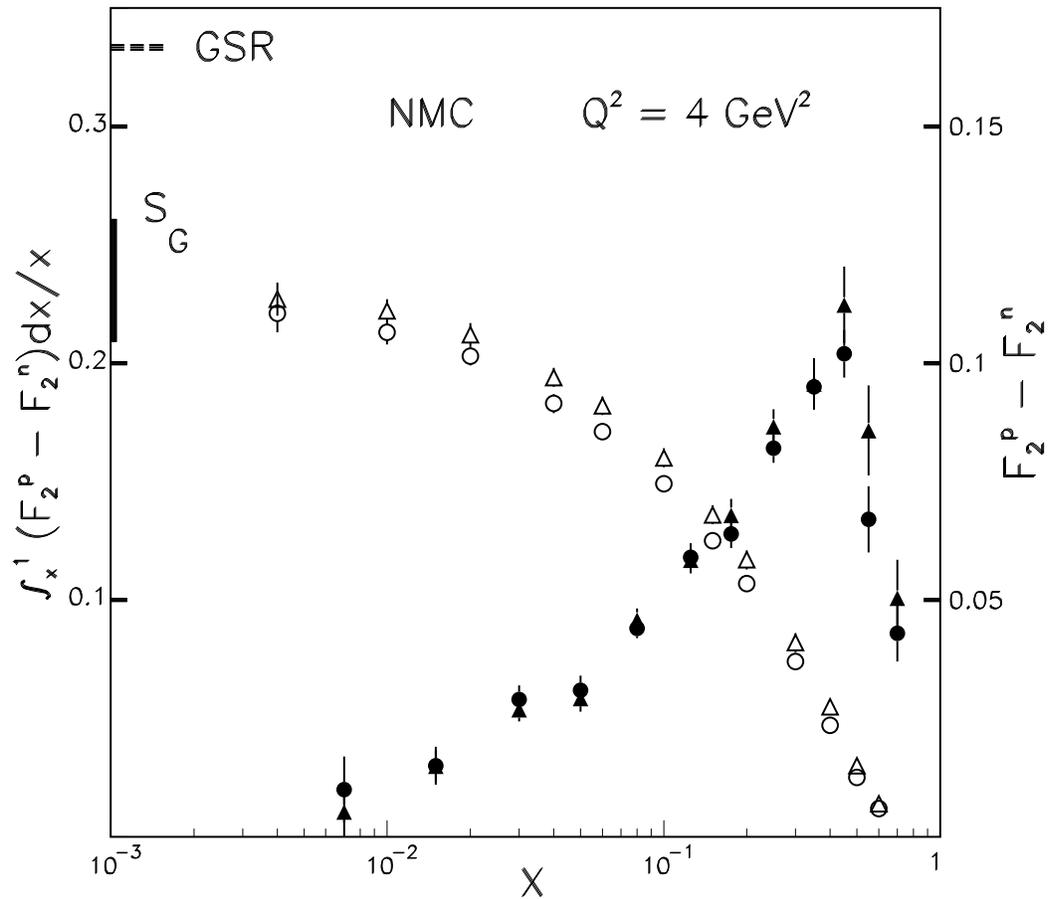,height=5.5in}
\caption{NMC measurements of $F^p_2 - F^n_2$ (solid data points)
and the Gottfried integral (open data points). The triangular data 
points correspond to results published in 1991~\cite{nmc91}, while
the circular data points represent a more recent analysis in 
1994~\cite{nmc94}. The extrapolated value of Gottfried integral ($S_G$)
and the expected GSR value are also indicated.}
\label{fig:3.1.1}
\end{figure}

The upgraded NMC detectors allowed a definitive study~\cite{nmc95} 
of the shadowing 
effect at low $x$. Moreover, they enabled a much more accurate 
determination of the Gottfried integral. Figure~\ref{fig:3.1.1} 
shows the $F^p_2 - F^n_2$ 
reported by the NMC in 1991~\cite{nmc91}, in which 
the smallest $x$ reached (0.004) was
significantly lower than in previous experiments. 
Taking advantage of their accurate measurements of the $F^n_2/F^p_2$ 
ratios, NMC used the following expression to evaluate $F^p_2 - F^n_2$, namely,
\begin{equation}
F^p_2 - F^n_2 = 2~F^d_2~(1 - F^n_2/F^p_2)/(1 + F^n_2/F^p_2).
\label{eq:3.1.1}
\end{equation}
The ratio $F^n_2/F^p_2 \equiv 2F^d_2/F^p_2 - 1$ was determined from NMC's
$F^d_2/F^p_2$ measurement, while $F^d_2$ was
taken from a fit to previous DIS experiments 
at SLAC~\cite{slac92}, Fermilab~\cite{chio79}, and
CERN~\cite{bcdms90a,emc90}. The value of the Gottfried integral 
for the measured region at $Q^2$ = 4 GeV$^2$ is 
$S_G(0.004 - 0.8) = 0.227 \pm 0.007(stat) \pm 0.014(syst)$.
The contribution to $S_G$ from $x > 0.8$ was estimated to be
$0.002 \pm 0.001$. Assuming that $F^p_2 - F^n_2$ at $x < 0.004$ behaves
as $ax^b$, NMC obtained $S_G(0 - 0.004) = 0.011 \pm 0.003$. Summing the 
contributions from all regions of $x$, NMC obtained 
$S_G = 0.240 \pm 0.016$,
which was significantly below 1/3. This represented the best evidence thus far
that the GSR was indeed violated.

In 1994, NMC reevaluated~\cite{nmc94} 
the Gottfried integral using a new $F^d_2$
parametrization from Ref.~\cite{nmc92} and newly determined 
values of $F^d_2/F^p_2$.
The new $F^d_2$ parametrization include data from 
SLAC~\cite{slac92}, BCDMS~\cite{bcdms90a}, as well as
NMC's own measurement~\cite{nmc92}. The new NMC 
results for $F^p_2 - F^n_2$ are shown 
in Figure~\ref{fig:3.1.1}. Note that the 1994 
values are slightly larger (smaller) than
the 1991 values at small (large) $x$. 
The new evaluation gave $S_G(0.004 - 0.8) = 0.221 \pm 0.008(stat) 
\pm 0.019(syst)$, and the Gottfried integral became
\begin{equation}
S_G = 0.235 \pm 0.026.
\label{eq:3.1.2}
\end{equation}
This  value is consistent with the earlier number. 
The new systematic error is larger than the old one, reducing somewhat 
the overall significance of the NMC measurement. Nevertheless, the violation 
of the GSR is established at a $4 \sigma$ level.

More recently, NMC published their final 
analysis of $F^d_2/F^p_2$~\cite{nmc97}
and $F^d_2$~\cite{nmc95a,nmc97a}. This analysis 
includes the 90 and 280 GeV data 
taken in 1986 and 1987, as well as the 1989 data at 120, 200 and 280 GeV.
The 1989 data were not used in the 
earlier evaluations~\cite{nmc91,nmc94} of the Gottfried
integral. Based on these new data, NMS reported a Gottfried integral 
in the interval
$0.004 < x < 0.8$ of $0.2281 \pm 0.0065(stat)$ 
at $Q^2$ = 4 GeV$^2$~\cite{nmc97}. 
This agrees within statistical errors with previous 
NMC results~\cite{nmc91,nmc94}.

QCD corrections to various parton-model sum rules have been reviewed
by Hinchliffe and Kwiatkowski~\cite{hinch96}. The $\alpha_s$ and
$\alpha_s^2$ corrections to the Gottfried sum have been 
calculated~\cite{ross79,kotaev96} and found to be very small (roughly
0.4 \% each at $Q^2$ = 4 GeV$^2$). Therefore, QCD corrections can not
account for the large violation of GSR found by the NMC. Although perturbative
QCD predicts a weak $Q^2$ dependence for the Gottfried sum, it has been 
suggested~\cite{forte94a, forte94b} that due to the non-perturbative origin of
the $\bar d$, $\bar u$ asymmetry the $Q^2$ dependence of 
the Gottfried sum will be anomalous between 1 and 5 GeV$^2$.
This interesting suggestion remains to be tested by DIS
experiments.\\

\noindent{\bf{3.2 E772 Drell-Yan Experiment}}

The main physics goal of the Fermilab experiment 772 was to 
examine the origin of the EMC effect. Among the many theoretical models 
which explain the EMC effect~\cite{gee95}, the 
pion-excess model~\cite{pionex83} predicted a large
enhancement of antiquark content due to the presence of additional meson
cloud in heavy nuclei. This prediction could be readily tested by measuring the
nuclear dependence of proton-induced DY cross sections. Using an 800 GeV proton 
beam, the DY cross section ratios of a number of nuclear targets 
($C, Ca, Fe, W$)
were measured~\cite{e772a} relative to deuterium 
with high statistics over the region
$0.04 < x_2 < 0.35$. While the enhancement of antiquark contents predicted by 
the pion-excess model was not observed, the E772 results
are in good agreement with the prediction of the 
rescaling model~\cite{close85}.

Information on the $\bar d/\bar u$ asymmetry has also been extracted from the
E772 data~\cite{e772b}. At $x_F > 0.1$, the dominant 
contribution to the proton-induced
DY cross section comes from the annihilation of $u$ quark in the projectile
with the $\bar u$ quark in the target nucleus. It follows that the DY cross 
section ratio of a non-isoscalar target (like $^1H, Fe, W$) over an isoscalar
target (like $^2H, ^{12}C, ^{40}Ca$) is given as
\begin{eqnarray}
R_A(x_2) \equiv \sigma_A(x_2)/\sigma_{IS}(x_2) \approx
1 + [(N-Z)/A] [(1 - \bar u(x_2)/\bar d (x_2))/(1 + \bar u(x_2)/\bar d(x_2))],
\label{eq:3.2.1}
\end{eqnarray}
where $x_2$ is the Bjorken-$x$ of the target partons, IS stands for isoscalar,
and $N, Z$ and $A$ refer to the non-isoscalar target. The $(N-Z)/A$ factor 
in Eq.~\ref{eq:3.2.1} shows that the largest sensitivity to the $\bar d/ \bar u$
could be obtained with a measurement of $\sigma(p+p)/\sigma(p+d)$.
Nevertheless, for $W$ target $(N-Z)/A = 0.183$ and the E772 
$\sigma_W/\sigma_{IS}$ data could still be used to study the $\bar d, \bar u$
asymmetry.

\begin{figure}
\center
\psfig{figure=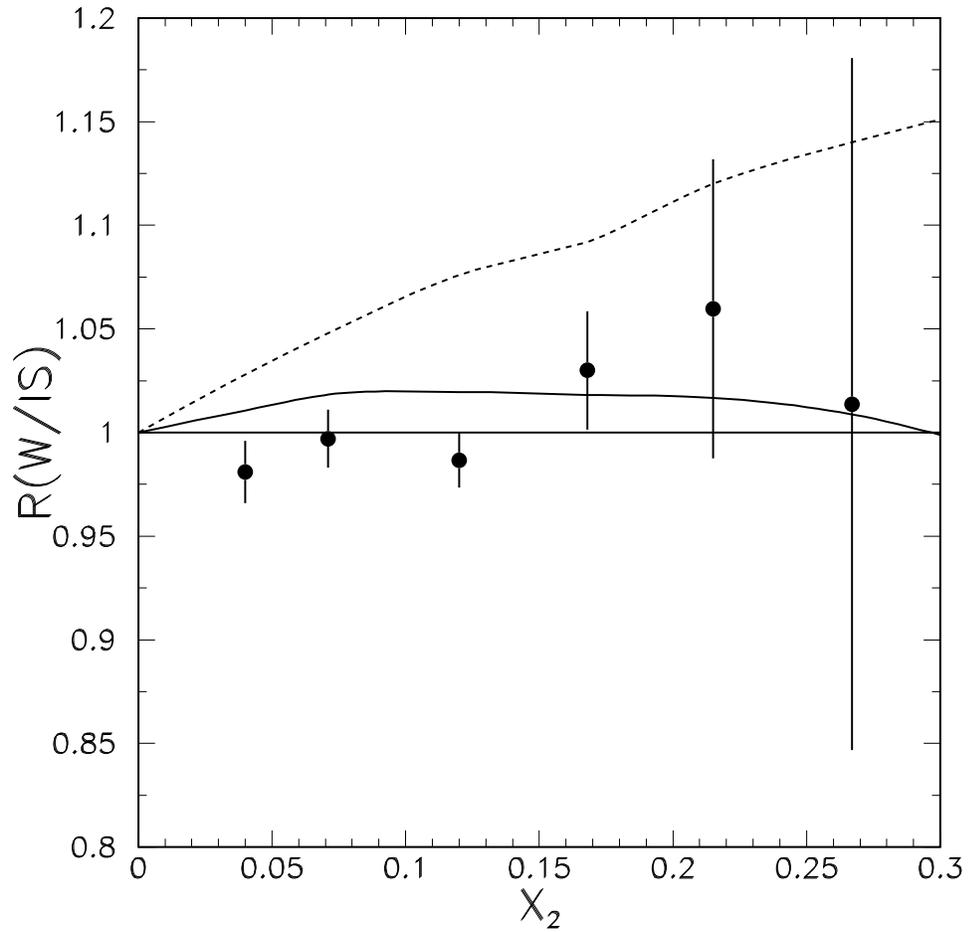,height=5.5in}
\caption{The ratios of $W$ over isoscalar targets DY cross sections
from E772~\cite{e772b}. The dashed curve is a calculation using the
$\bar d / \bar u$ asymmetric parton distributions suggested
in Ref.~\cite{es91}. The solid curve corresponds to a calculation
using the recent MRST distribution functions.}
\label{fig:3.2.1}
\end{figure}

Figure~\ref{fig:3.2.1} shows the E772 DY cross section ratios from 
the neutron-rich $W$ target over the isoscalar targets, $^2H$ and $^{12}C$.
Corrections have been applied to the two data points at $x_2 < 0.1$ to 
account for the nuclear shadowing effects in $C$ and $W$~\cite{e772b}. 
The E772 data 
provided some useful constraints on the $\bar d/\bar u$ asymmetry. 
In particular, some early parametrizations~\cite{es91,ehq92} 
of large $\bar d/\bar u$ 
asymmetry were ruled out. Despite the relatively large 
error bars, Figure~\ref{fig:3.2.1} shows $R > 1.0$ at $x_2 > 0.15$, 
suggesting that $\bar d > \bar u$ in this region. 

\begin{figure}
\center
\psfig{figure=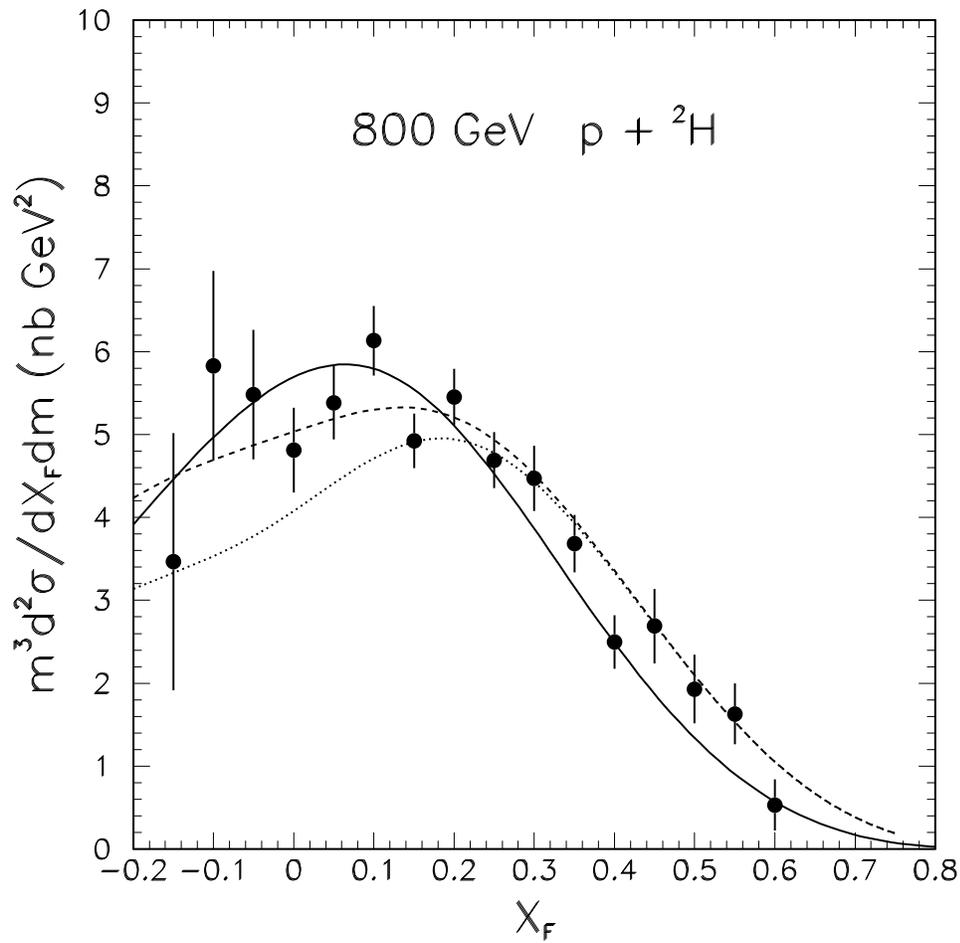,height=5.5in}
\caption{The $p + ^2H$ DY differential cross sections from E772~\cite{e772b}.
The dotted (dashed) curve is a calculation using the parton
distributions from Ref.~\cite{es91} with (without) 
$\bar d / \bar u$ asymmetry. The solid curve uses the MRST 
distribution functions. These are leading-order calculations normalized to 
the data.}
\label{fig:3.2.2}
\end{figure}

E772 collaboration also presented the DY differential cross sections for
$p+d$ at $\langle M_{\mu\mu}\rangle = 8.15$ GeV. As shown in Figure~\ref{fig:3.2.2},
the DY cross sections near $x_F = 0$ are sensitive to $\bar d/\bar u$
and the data disfavor a large $\bar d/\bar u$ asymmetry. Figure~\ref{fig:3.2.2}
also shows that a recent parton distribution function, 
MRST~\cite{mrst}, which has modest 
$\bar d/\bar u$ asymmetry, is capable of describing the $p+d$ differential 
cross sections well (see Section 3.6).

While the E772 data provide some useful constraints on the values 
of $\bar d/\bar u$, it is clear that a measurement of 
$\sigma_{DY}(p+d)/\sigma_{DY}(p+p)$ is highly desirable. The $(N-Z)/A$ 
factor is now equal to $-1$, indicating a large improvement in the 
sensitivity to $\bar d/\bar u$. Moreover, the uncertainty arising from
nuclear effects would be much reduced. It has been pointed out~\cite{kumano95}
that $\bar d/\bar u$ asymmetry in the nucleon could be significantly modified
in heavy nuclei through recombination of partons from different nucleons.
Clearly, an ideal approach is to first determine $\bar d/\bar u$ in the
nucleon before extracting the $\bar d/\bar u$ information in heavy nuclei.\\

\noindent{\bf{3.3 NA51 Drell-Yan Experiment}}

Following the suggestion of Ellis and Stirling~\cite{es91},
the NA51 collaboration at CERN carried out the first dedicated dimuon
production experiment to study the flavor structure of the 
nucleon sea~\cite{na51}.
Using a 450 GeV/c proton beam, roughly 2800 and 3000 dimuon events with 
$M_{\mu\mu} > 4.3$ GeV have been recorded, respectively, for $p+p$ and
$p+d$ interaction. The spectrometer setting covers the kinematic region
near $y = 0$. At $y = 0$, the asymmetry parameter, $A_{DY}$, is given as
\begin{eqnarray}
A_{DY} \equiv {\sigma^{pp} -\sigma^{pn}\over \sigma^{pp} + \sigma^{pn}}
\approx {(4\lambda_V - 1)(\lambda_s - 1) + (\lambda_V - 1)(\lambda_s - 1)\over
(4\lambda_V + 1)(\lambda_s + 1) + (\lambda_V + 1)(\lambda_s + 1)},
\label{eq:3.2.2}
\end{eqnarray}
where $\lambda_V = u_V/d_V$ and $\lambda_s = \bar u/\bar d$. 
In deriving Eq.~\ref{eq:3.2.2}, the negligible sea-sea annihilation 
was ignored and the validity of charge symmetry was assumed. At $x = 0.18$,
$\lambda_V \approx 2$ and according to Eq.~\ref{eq:3.2.2} $A_{DY} = 0.09$ for
a symmetric sea, $\lambda_s = 1$.
For an asymmetric $\bar d > \bar u$ sea, $A_{DY}$ would be less than 0.09.

From the DY cross section ratio, $\sigma^{pp}/\sigma^{pd}$, NA51
obtained $A_{DY} = -0.09 \pm 0.02 (stat) \pm 0.025 (syst)$. This then led
to a determination of $\bar u/\bar d = 0.51 \pm 0.04 (stat) \pm (syst)$
at $x = 0.18$ and $\langle M_{\mu\mu}\rangle = 5.22$ GeV. This 
important result established
the asymmetry of the quark sea at a single value of $x$. What remained
to be done was to map out the $x$-dependence of this asymmetry. This was
subsequently carried out by the Fermilab E866 
collaboration, as described in the next Section.\\

\noindent{\bf{3.4 E866 Drell-Yan Experiment}}

At Fermilab, a DY experiment (E866) aimed at a higher statistical
accuracy with a much wider kinematic coverage than the NA51 experiment was
recently completed~\cite{e866}. This experiment 
measured the DY muon pairs from 800
GeV proton interacting with liquid deuterium and hydrogen targets.
A proton beam with up to $2 \times 10^{12}$ 
protons per 20 s spill bombarded one 
of three identical 50.8 cm long cylindrical target flasks containing either
liquid hydrogen, liquid deuterium or vacuum. The targets alternated every 
few beam spills in order to minimize time-dependent systematic effects.
The dimuons accepted by a 3-dipole magnet spectrometer were detected by
four tracking stations. An integrated flux of $1.3 \times 10^{17}$ protons
was delivered for this measurement. 

\begin{figure}
\center
\psfig{figure=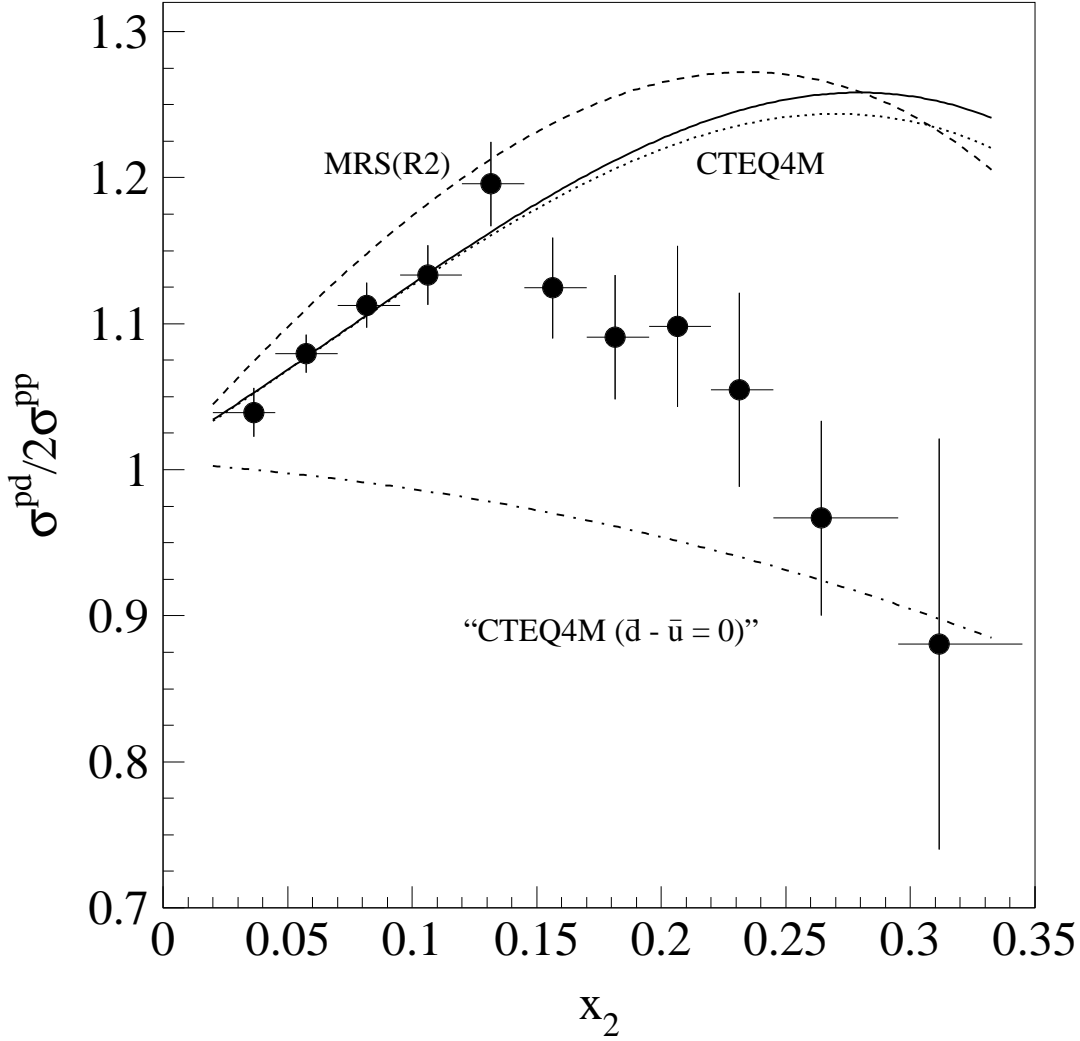,height=5.5in}
\caption{The ratio $\sigma^{pd}/2\sigma^{pp}$ of Drell-Yan cross
sections {\em vs.} $x_{2}$. The curves are next-to-leading order
calculations, weighted by acceptance, of the Drell-Yan cross section 
ratio using the CTEQ4M and MRS(R2) parton distributions. Also shown
is a leading-order calculation using CTEQ4M (dotted). In the lower
CTEQ4M curve $\bar{d} - \bar{u}$ has been arbitrarily set to 0 as
described in the text. The errors are statistical only.  There is
an additional 1\% systematic uncertainty common to all points.}
\label{fig:3.4.1}
\end{figure}

Over 330,000 DY events were recorded in E866, using three different 
spectrometer settings which covered the regions of low, intermediate and high 
mass muon pairs. The data presented here are from the high mass 
setting, with over 140,000 DY events. Analysis of the full data sets has been
completed and the results~\cite{towell} are in good 
qualitative agreement with the high-mass
data. The DY cross section ratio
per nucleon for $p + d$ to that for $p + p$ is 
shown in Figure~\ref{fig:3.4.1} as a function of 
$x_2$. The acceptance of the spectrometer
was largest for $x_F = x_1 - x_2 > 0$. In this kinematic regime the DY
cross section is dominated by the annihilation of a beam quark with a target
antiquark. To a very good approximation the DY cross section ratio
at positive $x_F$ is given as 
\begin{equation}
\sigma_{DY}(p+d)/2\sigma_{DY}(p+p) \simeq
(1+\bar d(x_2)/\bar u(x_2))/2.
\label{eq:3.3}
\end{equation}
In the case that $\bar d = \bar u$, the ratio is 1.
Figure~\ref{fig:3.4.1} shows that the DY cross section per nucleon for 
$p + d$ clearly exceeds $p + p$, and it indicates an excess of $\bar d$ 
with respect to $\bar u$ over an appreciable range in $x_2$.

Figure~\ref{fig:3.4.1} also compares the data with next-to-leading order (NLO)
calculations of the cross section ratio using the CTEQ4M~\cite{cteq}, 
MRS(R2)~\cite{mrs}, and a modified CTEQ4M parton
distributions.  The modified CTEQ4M parton distributions, in which the
$\bar{d}+\bar{u}$ parametrization was maintained but $\bar d$ was set 
identical to $\bar u$, were used to illustrate the cross section ratio
expected for a symmetric $\bar d/ \bar u$ sea. The E866 data clearly show 
that $\bar d \ne \bar u$. The data are in reasonable agreement with the 
unmodified CTEQ4M and the MRS(R2) predictions for $x_{2}<0.15$. Above 
$x_{2}=0.15$ the data lie well below the CTEQ4M and the MRS(R2) values.

\begin{figure}
\center
\psfig{figure=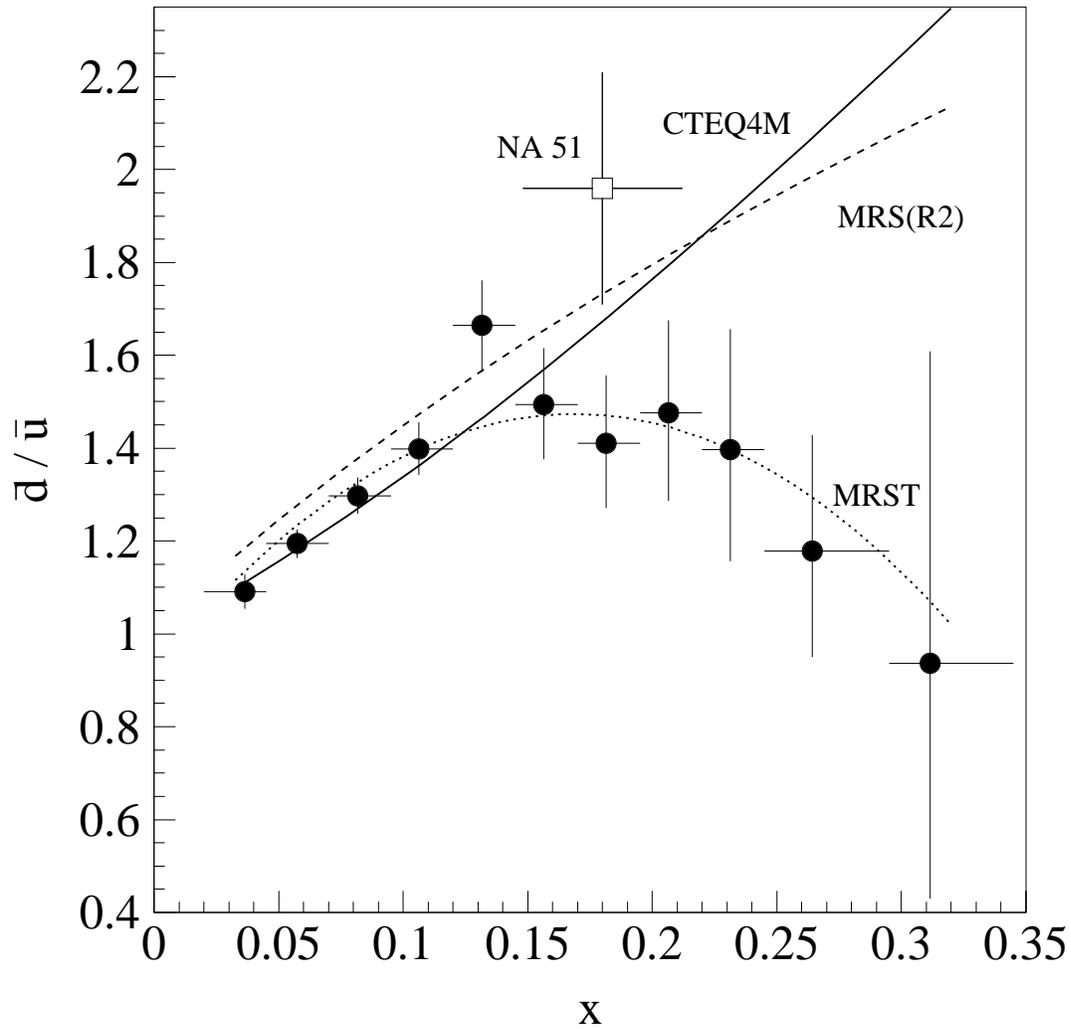,height=5.5in}
\vskip -0.3cm
  \caption{The ratio of $\bar{d}/\bar{u}$ in the proton as a function
  of $x_2$ extracted from the Fermilab E866 cross section ratio. The
  curves are from various parton distributions.  The error bars
  indicate statistical errors only.  An additional systematic
  uncertainty of $\pm0.032$ is not shown.  Also shown is the result
  from NA51, plotted as an open box.}
\label{fig:3.4.2}
\end{figure}

Using an iterative procedure described in \cite{e866,peng98}, values for 
$\bar d/ \bar u$ were extracted by the E866 collaboration at
$Q^2 = 54$ GeV$^2$ and shown in Figure~\ref{fig:3.4.2}.
At $x < 0.15$, $\bar d/\bar u$ increases linearly with $x$ and is in
good agreement with the CTEQ4M and MRS(R2) parametrization. 
However, a distinct feature of the data, not seen in either 
parametrization of the parton distributions, is the
rapid decrease towards unity of the $\bar{d}/\bar{u}$ ratio beyond
$x_{2}=0.2$\@. The result from NA51 is also shown in Figure~\ref{fig:3.4.2}
for comparison.

\begin{figure}
\center
\psfig{figure=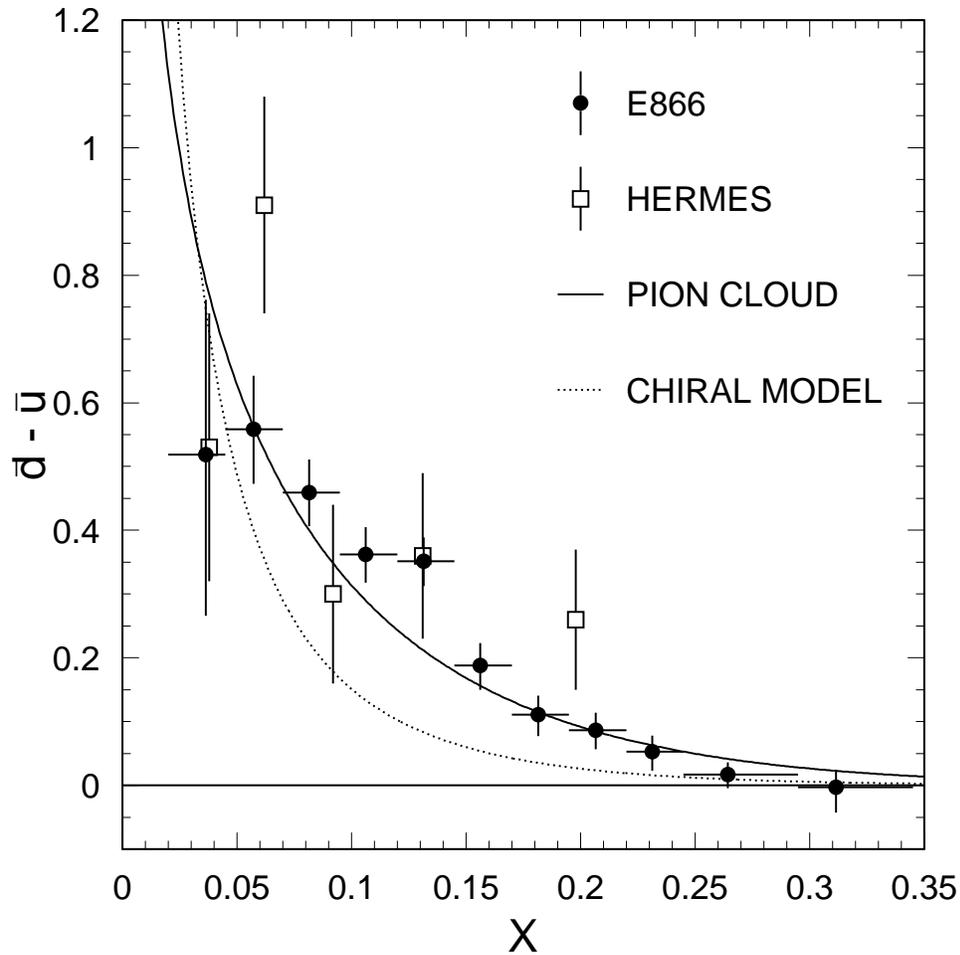,height=5.5in}
\caption{Comparison of the E866~\cite{e866} $\bar d - \bar u$ results at $Q^2$ =
54 GeV$^2$/c$^2$ with the predictions of pion-cloud and chiral models 
as described in the text. The data from HERMES~\cite{hermes98} are also shown.}
\label{fig:3.4.3}
\end{figure}

The $\bar d / \bar u$ ratios measured in E866, together with the
CTEQ4M values for $\bar d + \bar u$, were used to obtain
$\bar d - \bar u$ over the region $0.02 < x < 0.345$
(Figure~\ref{fig:3.4.3}). As a flavor non-singlet quantity,
$\bar d(x) - \bar u(x)$ is decoupled from the effects of the gluons and
is a direct measure of the contribution from non-perturbative processes, 
since perturbative processes cannot cause a significant $\bar d$, $\bar u$
difference. From the results shown in Figure~\ref{fig:3.4.3}, one can
obtain an independent determination~\cite{peng98} of the integral
of $\bar d(x) - \bar u(x)$ and compare it with the NMC result 
(Eq.~\ref{eq:3.1.2}). E866 obtains a value 
$\int_0^1 \left[\bar d(x) - \bar u(x)\right] dx = 0.100 \pm 0.007
\pm 0.017$, which is $2/3$ the value deduced by NMC~\cite{nmc94}.

\begin{figure}
\center
\psfig{figure=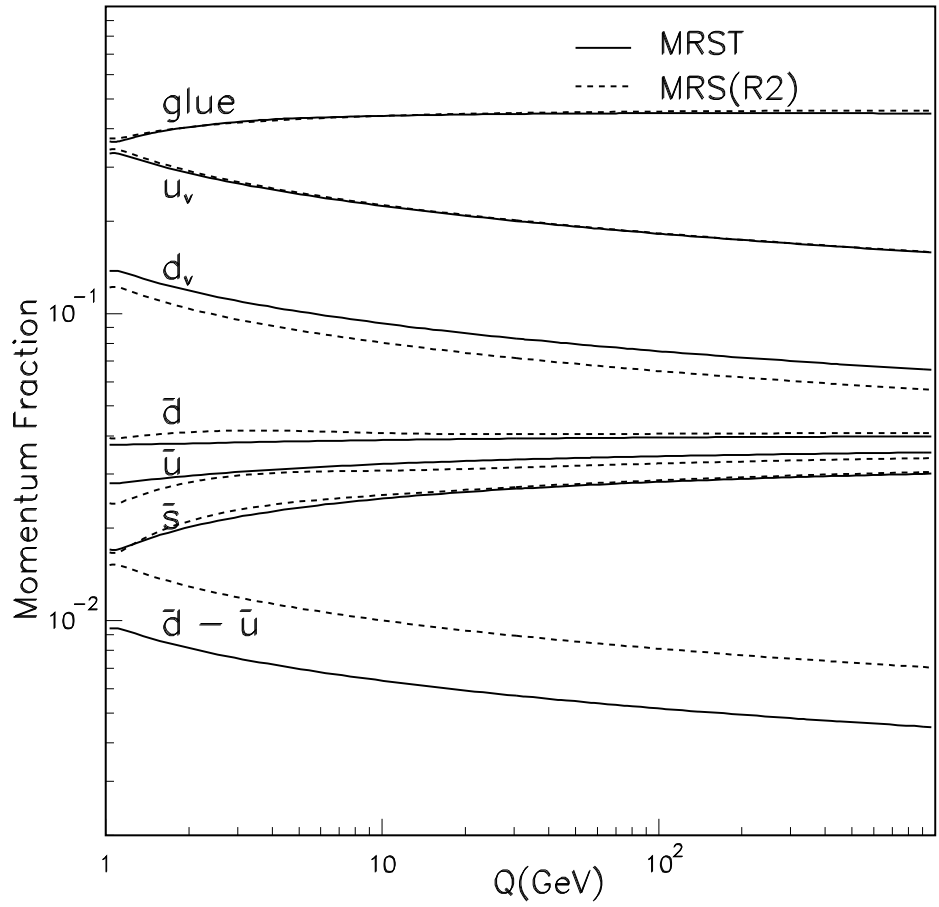,height=5.5in}
\caption{$Q$-dependence of the proton's momentum carried by
various partons calculated using the MRS(R2) and MRST parton
distribution functions.}
\label{fig:3.4.4}
\end{figure}

The E866 data also allow the first determination of the momentum fraction
carried by the difference of $\bar d$ and $\bar u$. 
One obtains 
$\int_{0.02}^{0.345} x
\left[\bar d(x) - \bar u(x)\right] dx = 0.0065 \pm 0.0010$ at 
$Q^2$ = 54 GeV$^2$.
If CTEQ4M is used to estimate
the contributions from the unmeasured $x$ regions, one finds that 
$\int_0^1 x \left[\bar d(x) - \bar u(x)\right] dx = 0.0075 \pm 0.0011$.
Unlike the integral of $\bar
d(x) - \bar u(x)$, the momentum integral is $Q^2$ dependent and
decreases as $Q^2$ increases. The $Q^2$ dependence of the momentum
fraction carried by various partons are shown in Figure~\ref{fig:3.4.4}.
The calculation uses both the MRS(R2) and the new MRST~\cite{mrst} 
parton distributions for comparison.\\

\noindent{\bf{3.5 HERMES Semi-Inclusive Experiment}}

It has been recognized for some time that semi-inclusive DIS could be
used to extract the flavor dependence of the valence
quark distributions~\cite{grz73}. From quark-parton model, the 
semi-inclusive cross
section, $\sigma^h_N$, for producing a hadron on a nucleon is given by
\begin{eqnarray}
{1 \over \sigma_N(x)} {d\sigma^h_N(x,z)\over dz} = 
{\sum_i e^2_i f_i(x) D^h_i(z)\over\sum_i e^2_i f_i(x)},
\label{eq:3.5.1}
\end{eqnarray}
where $D^h_i(z)$ is the fragmentation function signifying the probability for
a quark of flavor $i$ fragmenting into a hadron $h$ carrying
a fraction $z$ of the initial quark energy. $e_i$ and $f_i$ are the charge
and the distribution function of quark $i$, and $\sigma_N(x)$ is the
inclusive DIS cross section. Assuming
charge symmetry for the fragmentation functions and parton distribution
functions, one can derive the relationship
\begin{eqnarray}
{d_v(x)\over u_v(x)} ={4 R^\pi(x) + 1 \over 4 + R^\pi (x)}, 
\label{eq:3.5.2}
\end{eqnarray}
where
\begin{eqnarray}
R^\pi(x) = (d\sigma_n^{\pi^+}(x,z)/dz -d\sigma_n^{\pi^-}(x,z)/dz)/
(d\sigma_p^{\pi^+}(x,z)/dz -d\sigma_p^{\pi^-}(x,z)/dz).
\label{eq:3.5.3}
\end{eqnarray}
Based on a large number of semi-inclusive charged-hadron events in
muon DIS from hydrogen and deuterium targets, EMC extracted~\cite{ashman91}
the values
of $d_v(x)/u_v(x)$ over the range $0.028 < x < 0.66$. The EMC result agrees
with neutrino measurements~\cite{bebc84,cdhs84} using a different method, and 
it demonstrates the usefulness of such semi-inclusive measurements
for extracting information on valence quark distributions.

Soon after the report of GSR violation by the NMC, Levelt, Mulders and
Schreiber~\cite{levelt} pointed 
out that semi-inclusive DIS could also be used to
study the flavor dependence of sea quarks. In particular,
\begin{eqnarray}
{\bar d(x) - \bar u(x) \over u(x) - d(x)} = {J(z)[1 - r(x,z)] - [1 + r(x,z)]
\over J(z)[1 - r(x,z)] + [1 + r(x,z)]},
\label{eq:3.5.4}
\end{eqnarray}
where
\begin{eqnarray}
r(x,z) = {d\sigma_p^{\pi^-}(x,z)/dz - d\sigma_n^{\pi^-}(x,z)/dz \over 
d\sigma_p^{\pi^+}(x,z)/dz - d\sigma_n^{\pi^+}(x,z)/dz},~~ J(z) =
{3 \over 5}~ {1 + D_u^{\pi^-}(z)/D_u^{\pi^+}(z) \over 
1 - D_u^{\pi^-}(z)/D_u^{\pi^+}(z)}.
\label{eq:3.5.5}
\end{eqnarray}
Unlike the situation for $d_v(x)/u_v(x)$ which is completely independent of
the fragmentation functions $D_i^h$, Eq.~\ref{eq:3.5.4} shows that
fragmentation functions are needed to extract the values of 
$\bar d(x) - \bar u(x)$. 

Very recently, the HERMES collaboration~\cite{hermes98} at
DESY reported their measurements of charged hadrons produced in the
scattering of a 27.5 GeV positron beam on internal hydrogen, deuterium, and 
$^3$He target. The fragmentation functions $D_i^{\pi^\pm}(z)$
were extracted from the $^3$He data, while the hydrogen and deuterium
data allowed a determination of $r(x,z)$. The values of $(\bar d - \bar u)/
(u - d)$ show no $z$ dependence and are positive
over the region $0.02 < x < 0.3$, showing clearly an excess of $\bar d$ over
$\bar u$. Using the GRV94 LO~\cite{grv94} 
parametrization of $u(x) - d(x)$, the HERMES
collaboration obtained $\bar d(x) - \bar u(x)$  as shown 
in Figure~\ref{fig:3.4.3}. The integral of $\bar d - \bar u$ over the
measured $x$ region gives 
$\int_{0.02}^{0.3} \left[\bar d(x) - \bar u(x)\right] dx = 
0.107 \pm 0.021(stat) \pm 0.017(syst)$. The total integral over all
$x$ is extrapolated to be $0.16 \pm 0.03$, consistent with the result from
NMC. 
It is gratifying that the results from E866 and
HERMES are in rather good agreement, even though these two 
experiments use very different methods and cover very different 
$Q^2$ values ($Q^2 = 54$ GeV$^2$ in E866 and $Q^2 = 2.3$ GeV$^2$ in HERMES).

It should be mentioned that semi-inclusive DIS could 
be extended~\cite{frank89,close91} to
situations involving polarized lepton beam and polarized targets in order
to study the flavor dependence of the spin-dependent structure functions.
Both the Spin Muon Collaboration (SMC)~\cite{smc98} 
and the HERMES Collaboration~\cite{hermes99} have
reported the polarized valence quark distributions, $\Delta u_v(x)$ and
$\Delta d_v(x)$, and the non-strange sea-quark polarization, 
$\Delta \bar q(x)$.\\

\noindent{\bf{3.6 Impact on the Parton Distribution Functions}}

\begin{figure}
\center
\psfig{figure=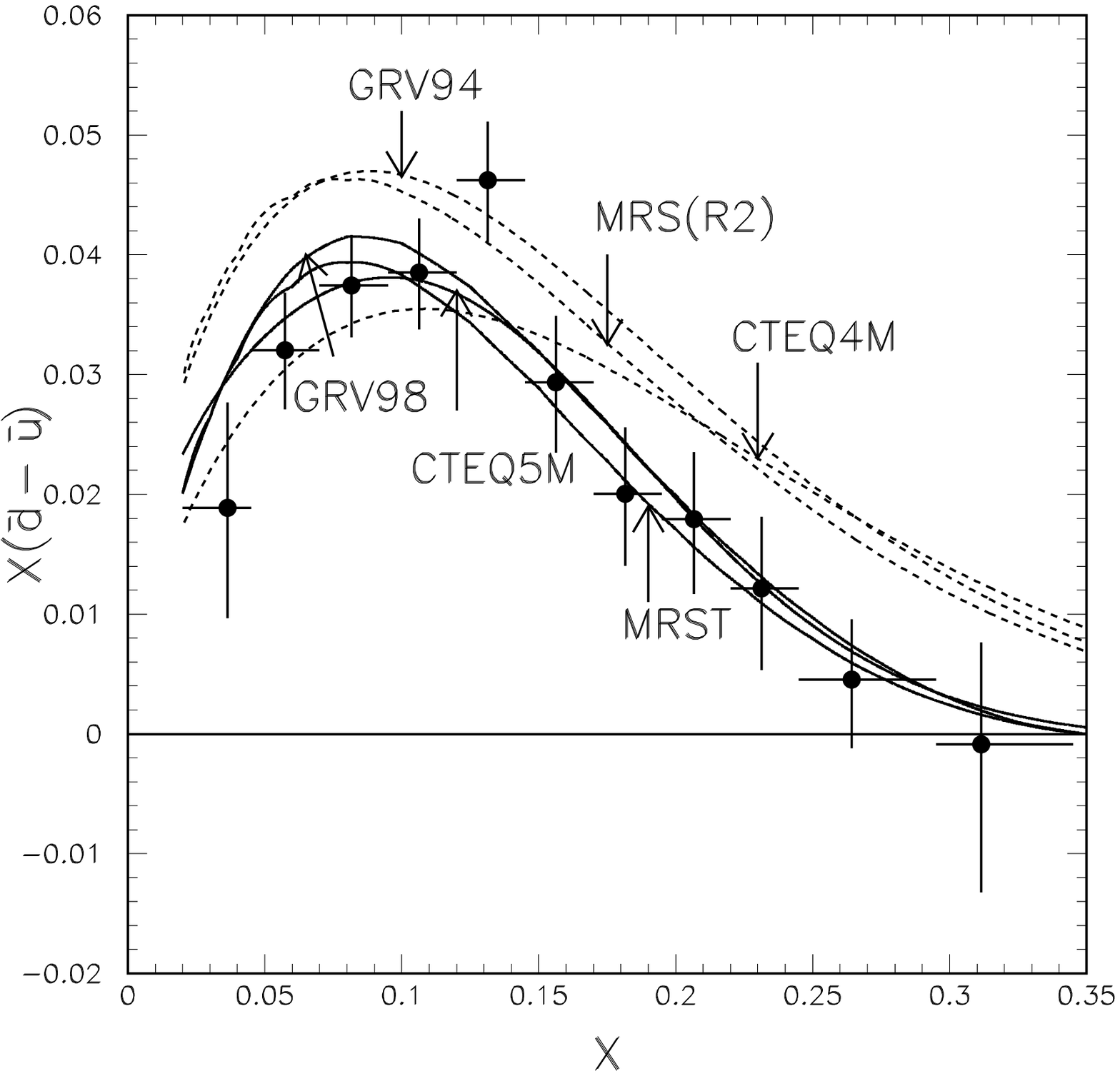,height=5.5in}
\caption{Comparison between the $x(\bar d - \bar u)$ results from E866
with the parametrizations of various parton distribution functions.
The dashed (solid) curves correspond to PDFs before (after) the E866
results were obtained.}
\label{fig:3.6.1}
\end{figure}

\begin{figure}
\center
\psfig{figure=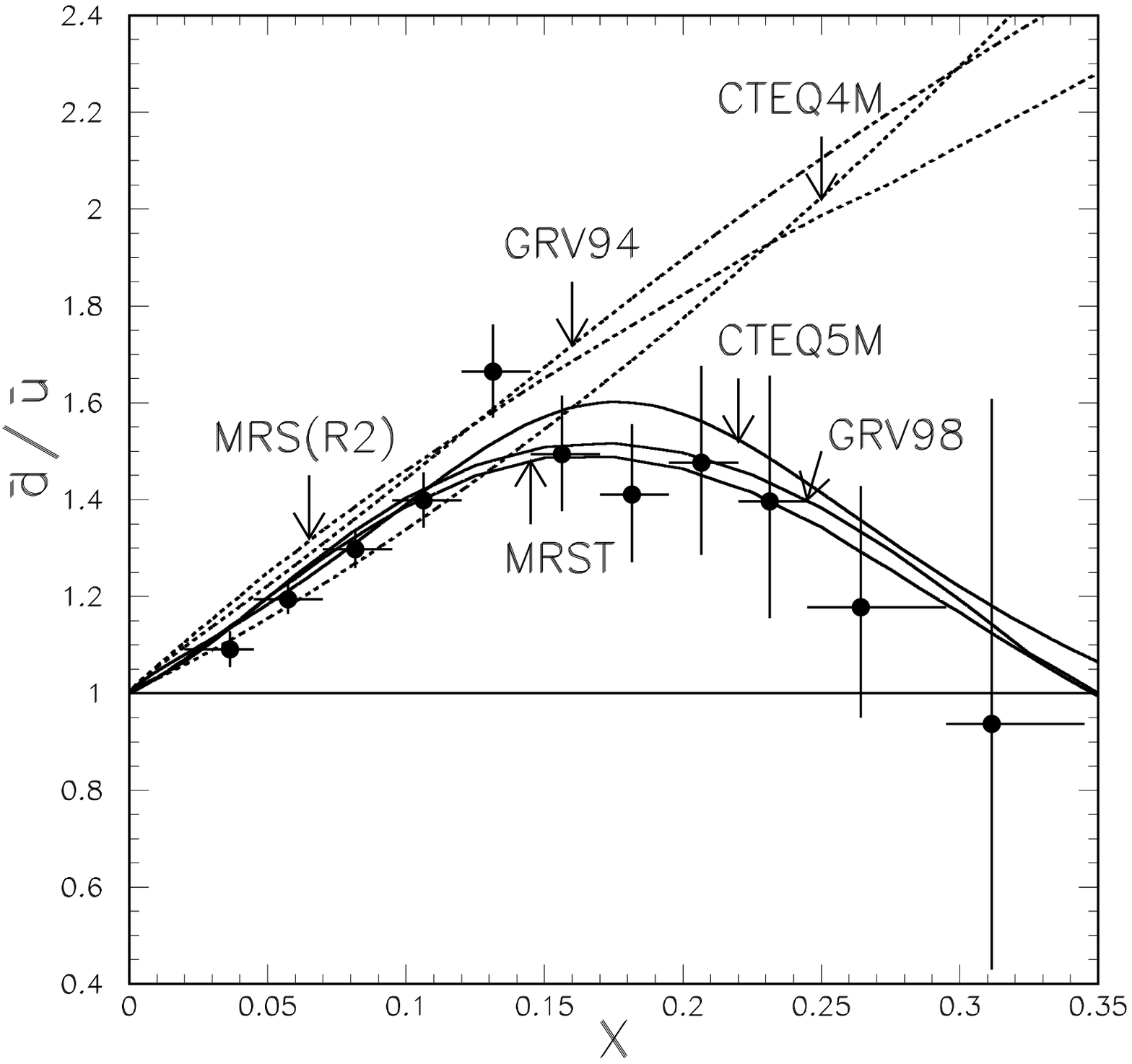,height=5.5in}
\caption{Comparison between the $\bar d / \bar u$ results from E866
with the parametrizations of various parton distribution functions.
The dashed (solid) curves correspond to PDFs before (after) the E866
results were obtained.}
\label{fig:3.6.2}
\end{figure}

Since evidences for a flavor asymmetric sea were reported by the NMC
and NA51, several groups~\cite{cteq,mrs,grv94} performing global analysis of 
parton distribution function (PDF) have all required $\bar d$ to be different
from $\bar u$. The NMC result constrained
the integral of the $\bar d - \bar u$ to be $0.149 \pm 0.039$, while the NA51
result requires $\bar d / \bar u$ to be $1.96 \pm 0.13$ at $x = 0.18$. Clearly,
the $x$ dependences of $\bar d - \bar u$ and $\bar d/ \bar u$ were
undetermined. Figures~\ref{fig:3.6.1} and ~\ref{fig:3.6.2} compare
the E866 measurements of $x(\bar d - \bar u)$ and $\bar d/ \bar u$
with the parametrizations of the MRS(R2)~\cite{mrs}, CTEQ4M~\cite{cteq}, 
and GRV94~\cite{grv94}, three
of the most frequently used PDF's prior to E866's measurement.

Recently, several PDF groups published new parametrizations taking
into account of new experimental information including the E866 data.
The parametrization of the $x$ dependence of $\bar d - \bar u$ is now
strongly constrained by the E866 and HERMES data. In particular,
$\bar d(x) - \bar u(x)$ or $\bar d(x)/\bar u(x)$ are parametrized
as follows;

\noindent MRST~\cite{mrst}:
\begin{eqnarray}
\bar d(x) - \bar u(x) = 1.29 x^{0.183} (1-x)^{9.808} (1+9.987x-33.34x^2),
~Q_0^2 = 1~GeV^2,  \nonumber
\end{eqnarray}
GRV98~\cite{grv98}:
\begin{eqnarray}
\bar d(x) - \bar u(x) = 0.20 x^{-0.57} (1-x)^{12.4} (1-13.3x^{0.5}+60.0x),
~Q_0^2 = 0.4~GeV^2, \nonumber
\end{eqnarray}
CTEQ5M~\cite{cteq5, lai99a}:
\begin{eqnarray}
\bar d(x) / \bar u(x) = 1 - 1.095 x + 3.159 x^{0.5}/[1+((x-0.188)/0.116)^2],
~Q_0^2 = 1~GeV^2.
\label{eq:3.6.1}
\end{eqnarray}
As shown in Figure~\ref{fig:3.6.2}, these new 
parametrizations give significantly different shape for $\bar d/ \bar u$
at $x > 0.15$ compared to previous parametrizations.
Table~\ref{tab:3.6.1} also lists the values of $\bar d - \bar u$ integral
from various recent PDF's.

\begin{figure}
\center
\psfig{figure=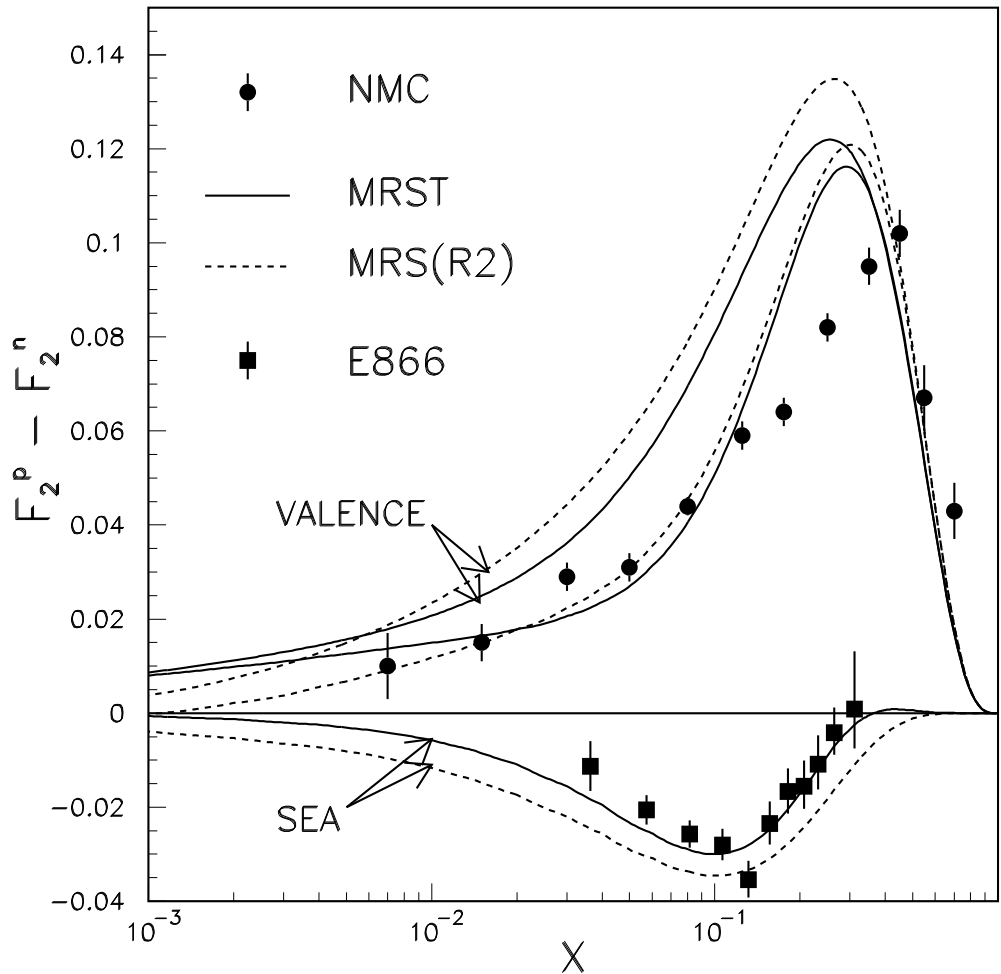,height=5.5in}
\caption{$F^p_2 - F^n_2$ as measured by NMC at $Q^2$ = 4 GeV$^2$ compared with
predictions based on the MRS(R2) and MRST parametrizations. Also
shown are the E866 results, evolved to $Q^2$ = 4 GeV$^2$, for 
the sea-quark contribution to $F^p_2
- F^n_2$. For each prediction, the top (bottom) curve is the valence
(sea) contribution and the middle curve is the sum of the two.}
\label{fig:3.6.3}
\end{figure}

It is interesting to note that the E866 data
also affect the parametrization of the valence-quark distributions.
Figure~\ref{fig:3.6.3} shows the NMC data for $F_2^p
- F_2^n$ at $Q^2$ = 4 GeV$^2$, together with the fits of MRS(R2) and
MRST. It is instructive to
decompose $F_2^p(x) - F_2^n(x)$ into contributions from 
valence and sea quarks:
\begin{eqnarray}
F_2^p(x) -F_2^n(x) = {1 \over 3} x \left[u_v(x) - d_v(x)\right] + {2 \over
3} x \left[\bar u(x) - \bar d(x)\right].
\label{eq:3.6.2}
\end{eqnarray}
As shown in Figure~\ref{fig:3.6.3}, the E866 data
provide a direct determination of the sea-quark contribution to $F_2^p
- F_2^n$. In order to preserve the fit to $F_2^p - F_2^n$, the MRST's
parametrization for the valence-quark distributions, $u_v - d_v$,
is significantly lowered in the region $x > 0.01$. Indeed, one of the major new
features of MRST is that $d_v$ is now significantly higher than before at
$x > 0.01$. Although the authors of MRST attribute this to the new 
$W$-asymmetry data from CDF~\cite{cdf98} and 
the new NMC results on $F_2^d/F_2^p$~\cite{nmc97}, it 
appears that the new information on $\bar d(x) - \bar u(x)$ has a direct
impact on the valence-quark distributions too.

\begin{figure}
\center
\psfig{figure=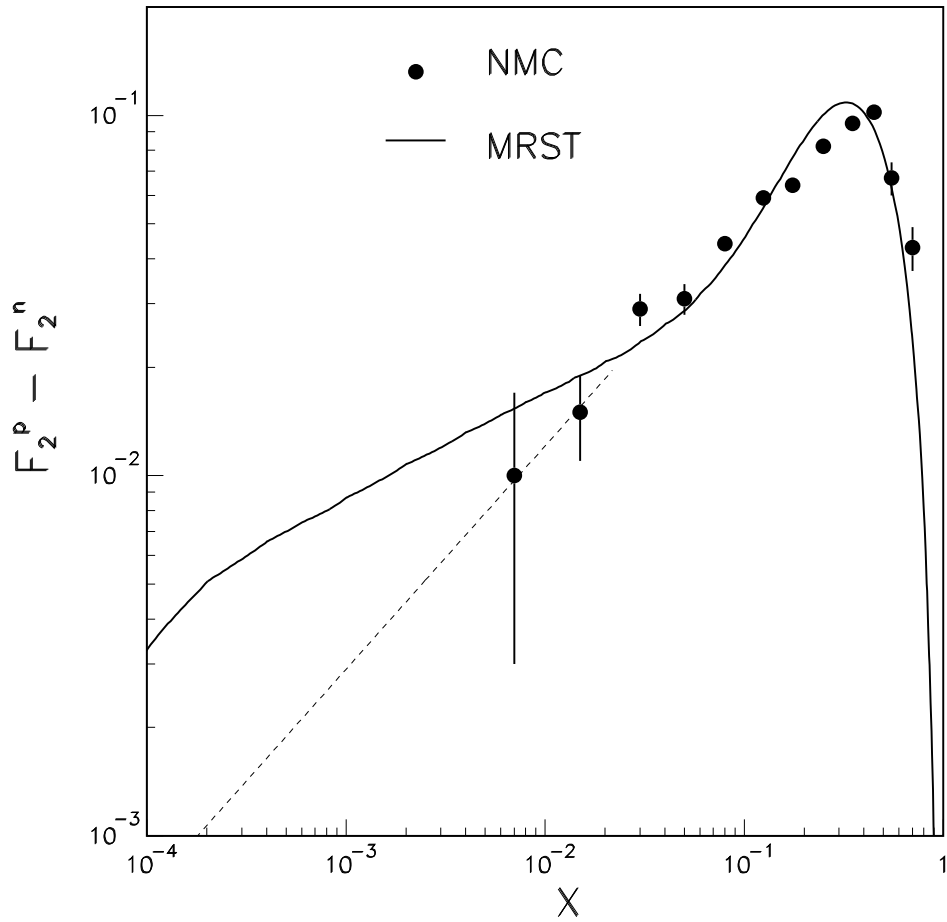,height=5.5in}
\caption{$F^p_2 - F^n_2$ as measured by NMC at $Q^2$ = 4 GeV$^2$ compared with
parametrization of MRST. The dashed curve corresponds to $0.21 x^{0.62}$,
a parametrization assumed by the NMC for the unmeasured small-$x$ region
when the Gottfried integral was evaluated.}
\label{fig:3.6.4}
\end{figure}

Another implication of the E866 data is on the behavior of
$F_2^p - F_2^n$ at small $x$. In order to satisfy the constraint
$\int_0^1 [u_v(x) - d_v(x)] dx = 1$, the MRST values of $u_v(x) - d_v(x)$
at $x < 0.01$ are now much larger than in MRS(R2), since
$u_v(x) -d_v(x)$ at $x > 0.01$ are smaller than before. 
As a consequence, $F_2^p - F_2^n$
is increased at small $x$ and MRST predicts a large contribution to the
Gottfried integral from the small-$x$ ($x < 0.004$) region, as shown in
Figure~\ref{fig:3.6.4}. If the MRST
parametrization for $F_2^p - F_2^n$ at $x < 0.004$ were used, NMC
would have deduced a value of 0.252 for
the Gottfried integral, which would imply a value of 0.122 for 
the $\bar d - \bar u$ integral. This would bring better agreement 
between the E866 and the NMC results on the $\bar d - \bar u$ integral.\\

\noindent{\bf{3.7 Comparison Between Various Measurements}}

Are the various measurements of sea quark flavor asymmetry consistent among 
them? In particular,
is the E866 result consistent with the earlier E772 and the NA51
DY experiments, and with the HERMES and NMC DIS measurements? To address 
this question, we first compare E866 with E772, both of which are 
DY experiments using 800 GeV proton beam with essentially the
same spectrometer. Figures~\ref{fig:3.2.1} and~\ref{fig:3.2.2} show that
the MRST parton distributions, which determined $\bar d - \bar u$ based on 
the E866 data, can also describe the E772 data very well, and we conclude that
the E866 and E772 results are consistent.

Although both NA51 and E866 measured $\sigma (p+d)/ 2 \sigma (p+p)$ to extract
the values of $\bar d/ \bar u$, some notable differences exist. As mentioned
earlier, NA51 measured the ratio at a single value of $x_2$ ($x_2=0.18)$ near
$x_F \approx 0$ using 450 GeV proton beam, while E866 used 800 GeV
proton beam to cover a broader range of $x_2$ at $x_F > 0$.
It is instructive to compare the NA51 result at $x_2 = 0.18$ with
the E866 data at $x_2 = 0.182$. Table~\ref{tab:3.7.1} lists the
kinematic variables and physics quantities derived from these two 
data points. It is interesting to note that the values of 
$\sigma (p+d)/ 2 \sigma (p+p)$ at $x_2 = 0.18$ are actually 
very similar for NA51 and E866, even though the derived values 
for $\bar d/ \bar u$ differ significantly. This reflects the difference
in $x_F$ for both experiments, making the values of $\bar d/ \bar u$
extracted from $\sigma (p+d)/ 2 \sigma (p+p)$ different. The other difference
is $Q^2$, being a factor of 3.6 higher for E866. Using MRST and CTEQ5M 
to estimate
the $Q^2$ dependence of $\bar d/ \bar u$, we find that the NA51 value
of $\bar d/ \bar u$ is reduced by $\approx$ 3 \% going 
from $Q^2$ = 27.2 GeV$^2$ to $Q^2$ = 98.0 GeV$^2$. This brings slightly better
agreement bewteen NA51 and E866.

The methods used by HERMES and E866 to determine $\bar d - \bar u$ are
totally different, and it is reassuring that the results came out
alike, as shown in Figure~\ref{fig:3.4.3}. The $\bar d - \bar u$ values
from HERMES are in general somewhat larger than those of E866. 
At a relatively low mean $Q^2$ of 2.3 GeV$^2$, the HERMES experiment
could be subject to high-twist 
effect~\cite{ball94}. Additional data from HERMES
are expected to improve the statistical accuracy.

The comparison between E866 and NMC in terms of the integral of $\bar d -
\bar u$ has been discussed earlier. A possible origin for the apparent
differences of the integral was also discussed in Section 3.6.\\

\noindent{\bf{4. ORIGINS OF THE $\bar d/ \bar u$ ASYMMETRY}}

\noindent{\bf{4.1 Pauli Blocking}}

The earliest experiments indicated that the value of the Gottfried
integral might be less than 1/3, leading to speculation regarding the origin
of this reduction. Field and Feynman suggested~\cite{field77} 
that it could be due to Pauli blocking in so far as $u \bar u$ pairs 
would be suppressed relative to $d \bar d$ pairs
because of the presence of two $u$-quarks in proton as compared to a single
$d$-quark.  Ross and Sachrajda~\cite{ross79} 
questioned that this effect would be
appreciable because of the large phase-space available to the created
$q \bar q$ pairs. They also showed that perturbative QCD would not 
produce a $\bar d$, $\bar u$ asymmetry. Steffens and Thomas~\cite{stef97} 
recently looked into this issue,
explicitly examining the consequences of  Pauli blocking. They
similarly concluded that the blocking effects were small, 
particularly when the antiquark is in a virtual meson.

The small $d, u$ mass difference (actually, $m_d > m_u$) of~ 2 to 4 MeV
compared to the nucleon confinement scale of 200 MeV/c does not permit any
appreciable difference in their relative production by gluons. At any rate,
one observes a surplus of $\bar d$ which is the heavier of the two species.
As pointed out above, blocking effects arising from the Pauli exclusion
principle should also have little effect. Thus another, presumably
non-perturbative, mechanism must be found to account for the large measured
$\bar d, \bar u$ asymmetry.\\

\noindent{\bf{4.2 Meson-Cloud Models}}

A natural origin for this flavor asymmetry is the virtual states of
the proton containing isovector mesons. This point appears to have first
been made by Thomas in a publication ~\cite{thomas83} 
treating SU(3) symmetry breaking in
the nucleon sea. Sullivan~\cite{sullivan} 
had earlier shown that virtual meson-baryon
states directly contribute to the nucleon''s structure function. A large
number of authors have contributed to calculating the asymmetry from 
this perspective, so recent reviews~\cite{kumano98,speth98} should
be consulted for a complete list of contributions.

Conservation of electric charge and isospin naturally creates a flavor
asymmetry when isovector mesons are involved. For example the Fock
decomposition of the proton into its $\pi N$ components yields 
\begin{eqnarray}
|p\rangle \to \sqrt{1 \over 3}~ |p_0 \pi^0 \rangle + \sqrt{2 \over 3}~
|n_0 \pi^+\rangle,
\label{eq:4.2.1}
\end{eqnarray}
where $p_0$ and $n_0$ are regarded as proton and neutron states with 
symmetric seas. The $\bar d/ \bar u$ ratio
for the configurations shown in Eq.~\ref{eq:4.2.1} is 5!  Thus the 
presence of such
virtual states can readily generate a large flavor asymmetry. Allowing 
for the possibility of $\pi \Delta$ configurations it is easy to show that the
integrated  $\bar d, \bar u$ difference in the proton is given by
\begin{eqnarray}
\int_0^1 [\bar d(x) - \bar u(x)] dx = {1 \over 3} (2a-b),
\label{eq:4.2.2}
\end{eqnarray}
where $a(b)$ is the probability of $\pi N (\pi \Delta)$ 
component in the proton. The value
extracted from E866 for the above integral is
$0.100 \pm 0.018$, leading to $a = 0.2 \pm 0.036$
if one accepts $b \approx a/2$ as found in many calculations. 

Many attempts~\cite{henley,kumano,signal,hwang,szczurek,koepf,wally98} 
have been made to calculate the flavor asymmetry due to
isovector mesons. Most start with the following convolution expressions:

\begin{eqnarray}
|p\rangle = a_0|p_0\rangle + \sum_{MB} a^p_{MB} |M\rangle |B\rangle,
\label{eq:4.2.3}
\end{eqnarray}
\begin{eqnarray}
x\bar q_p(x,Q^2) = \sum_{MB} a^p_{MB} \int_x^1 dy~ f_{MB}(y)~ {x \over y}~
\bar q_M({x \over y},Q^2),
\label{eq:4.2.4}
\end{eqnarray}
where
\begin{eqnarray}
f_{MB}(y) = {g^2_{MpB} \over 16\pi^2} y \int_{-\infty}^{t_{min}} dt 
{F(t,m_p,m_B) \over (t - m^2_M)^2} F^2_{MpB}(t,\Lambda),~~ 
t_{min} = m^2_p y - m^2_B {y \over 1-y}.
\label{eq:4.2.5}
\end{eqnarray}
In the above expressions $x$ is the fraction of proton''s momentum carried by
the antiquark, and
$y$ is fraction carried by the meson ($M$). The meson-proton-baryon 
couplings are
characterized by coupling constants $g_{MpB}$, and form 
factors $F_{MpB}(t,\Lambda)$ where $\Lambda$ is a cutoff parameter. 
$F(t,m_p,m_B)$ is a kinematic factor depending
on whether $B$ is in the baryon octet or decuplet. 
As pions are the only mesons usually considered and the baryons are usually
restricted to nucleons and deltas, the coupling constants are well known
and the partonic structure of the pion, $\bar q_\pi (x,Q^2)$, is fixed 
by measurement of the
DY process using high energy pion beams. The only uncertainties are
the form factors $F_{\pi pN}(t)$ and $F_{\pi p \Delta}(t)$. One 
attempts to determine these form
factors by using~\cite{holt96,niko99} the measured yields from 
a variety of high energy hadronic reactions at small 
$p_T$ such as $p + p \to n + X$ and $p + p \to \Delta^{++} + X$.
Even though there is a sizable amount of available data, employing such a
procedure does not produce a precise result. The cutoff parameters used 
in the extracted 
form factors are of the order of 1 GeV  but the uncertainties in their values
produce factors of two in the predicted antiquark content of the nucleon.

Even though calculated value of the integral of $\bar d(x) - \bar u(x)$
is often in agreement with
experiment, it is more difficult to achieve a quantitative fit to the
measured $x$ dependence of the difference~\cite{peng98}. 
Figure~\ref{fig:3.4.3} compares $\bar d(x) - \bar u(x)$ from E866
with a pion-cloud model calculation, following the procedure detailed
by Kumano~\cite{kumano}. A dipole form, with $\Lambda = 1.0$ GeV for 
the $\pi N N$ form factor and $\Lambda = 0.8$ GeV for 
the $\pi N \Delta$ form factor,
was used. Calculations of $\bar d(x)/ \bar u(x)$  
are even more unsuccessful, as
knowledge of the $x$ dependence of the symmetric sea is required in this
instance~\cite{peng98}. 

\begin{figure}
\center
\psfig{figure=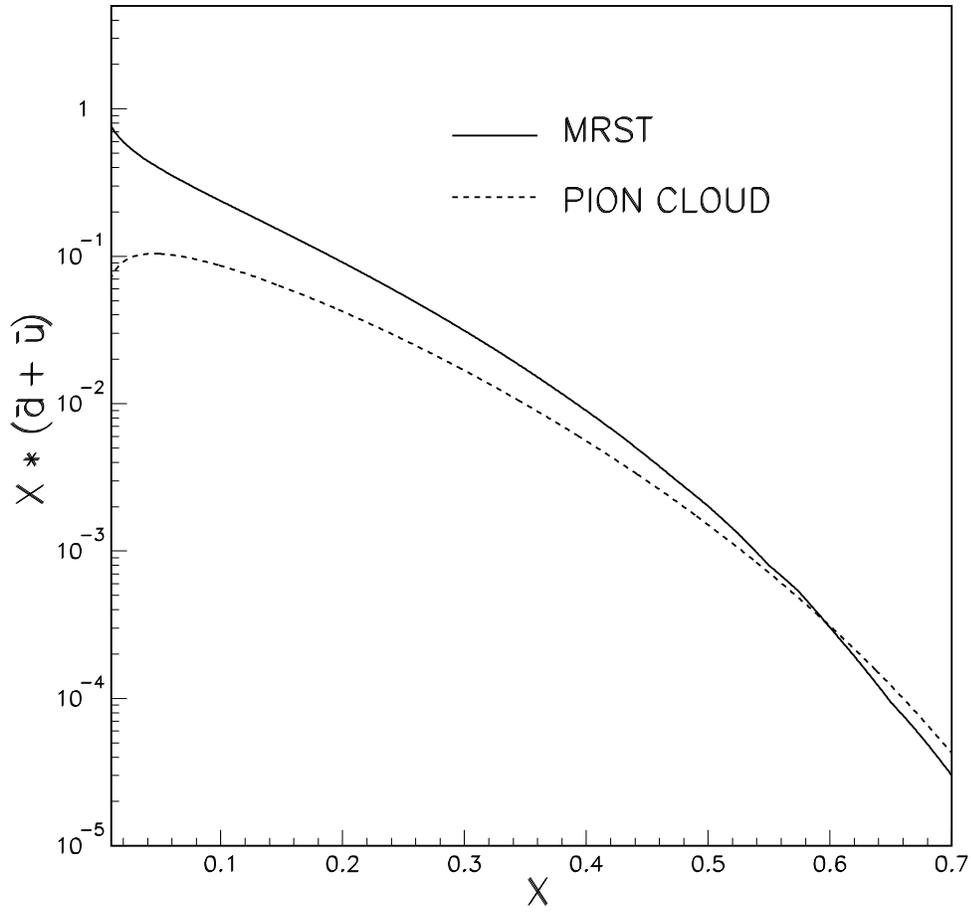,height=5.5in}
\caption{Comaprison of $x (\bar d + \bar u)$ obtained from the valence
quarks in the pion cloud with 
the parametrization of the MRST parton distribution functions.}
\label{fig:4.2.1}
\end{figure}

It is instructive to compare the pion-model prediction with
the current PDF parametrization of $x(\bar d + \bar u)$.
Figure~\ref{fig:4.2.1} shows that at small $x$ ($x < 0.1)$ 
the valence quarks in the
pion cloud account for less than $1/3$ of the $\bar d + \bar u$ content in the
proton. In contrast, at large $x$ ($x > 0.5$) the pion model would 
attribute all of $\bar d + \bar u$ to the pion cloud.

More recent attempts~\cite{holt96,niko99} to calculate the asymmetry due to
isovector mesons
find a smaller $\pi \Delta$ component in the nucleon than is presented
following Eq.~\ref{eq:4.2.2}. The $\pi N$ component is still 
about 20\% while the $\pi \Delta$ piece is
down around 6\%. As a result the integrated asymmetry is too large,
typically 0.16. However considerable progress has been made by Reggeizing
the virtual mesons~\cite{niko99}. This procedure has two interesting 
consequences, first it shows that the principal contributions to the
asymmetry come from virtual pions and secondly it reduces the 
contributions from these pions at large $x$, allowing the ratio 
$\bar d(x)/\bar u(x)$ to approach 1 for $x > 0.3$. Another recent 
work~\cite{magnin99} describes meson-baryon fluctuation
through gluon splitting and a phenomenological 
recombination mechanism, and it also reproduces the
$x$ dependence of $\bar d/ \bar u$ well.

Recently, the flavor asymmetry of the nucelon sea was computed 
in the large-$N_c$ limit, where the nucleon is described as a 
soliton of an effective chiral theory~\cite{waka98, poby99}.
In this chiral quark-soliton model, the flavor non-singlet 
distribution, $\bar u(x) - \bar d(x)$, appears in the next-to-leading
order of the $1/N_c$ expansion~\cite{diak96, diak97}. The E866 
$\bar d(x) - \bar u(x)$ data were shown to be well described by 
this model~\cite{poby99}.\\

\noindent{\bf{4.3 Chiral Models}}

An alternative 
approach ~\cite{ehq92,li95,li97,weise98,szc96,song97,ohlsson99} also 
employing virtual pions to produce 
the $\bar d, \bar u$ asymmetry uses 
constituent quarks and pions as the relevant degrees
of freedom. Such models are usually referred to as 
chiral models~\cite{georgi84}. 
In this model, a portion of the sea comes from the couplings of Goldstone
bosons to the constituent quarks, such as $u \to d \pi^+$ and $d \to u
\pi^-$. The excess of $\bar d$ over $\bar u$ is then simply due to the
additional up valence quark in the proton. 

The chiral models have 
less dynamical freedom, and the meson-baryon expansion can be
shown to predict  $\bar d(x)/ \bar u(x) < 11/7$ independent of $x$ or $Q^2$. 
This limit is exceeded
for  $0.12 < x < 0.2$ as shown in Figure~\ref{fig:3.4.2}.
The limit on the maximum value of the ratio can be traced to the fact that
there is no mechanism to suppress virtual $\pi \Delta$-like 
configurations which are
seen in Eq.~\ref{eq:4.2.2} to reduce the asymmetry.  
Some extensions of the simple chiral
quark model can obtain agreement with certain features of the observed
asymmetry but only with the introduction of additional parameters. A
calculation~\cite{peng98, szc96} of 
$\bar d(x) - \bar u(x)$ (Figure~\ref{fig:3.4.3}) 
yielded too soft a distribution relative to experiment, 
indicating~\cite{peng98} that more dynamics must be included in
the chiral quark model if it is to produce more than rough 
qualitative agreement with the observed asymmetry in the up, down 
sea of the nucleon.\\

\noindent{\bf{4.4 Instanton Models}}

Instantons have been known as theoretical constructs since the
seventies~\cite{bel75,hooft76,shu98}. 
They represent non-perturbative fluctuations of the gauge fields
that induce transitions between degenerate ground states of different
topology. In the case of QCD, the collision between a quark and an
instanton flips the helicity of the quark while creating a $q \bar q$
pair of
different flavor. Thus, interaction between a $u$ quark and an instanton
results in a $u$ quark of opposite helicity and either a $d \bar d$ 
or $s \bar s$ pair. 
Such a model has the possibility of accounting for both the flavor
asymmetry and the ``spin crisis"~\cite{forte89,forte91}. 
However, the model has proven difficult to exploit for
this purpose. There is only one case~\cite{inst93} of its being 
employed to explain these
anomalous effects. In the case of the $\bar d, \bar u$ flavor 
asymmetry the authors of ref.~\cite{inst93} fit
the instanton parameters to reproduce the violation
of the GSR observed by NMC. 
The prediction~\cite{inst93} at large $x$, $\bar d(x) / \bar u(x) \to 4$, 
is grossly violated by experiment (see Figure~\ref{fig:3.4.2}). 
Thus, it appears that while instantons have the
possibility for accounting for flavor and spin anomalies, the approach is not
yet sufficiently developed for a direct comparison. The final state created
via an instanton collision is quite similar to that created via the
emission of meson in the chiral model.\\

\noindent{\bf{4.5 Charge Symmetry Violation}}

Charge symmetry is believed to be well respected in strong interaction.
Extensive experimental searches for charge symmetry violation (CSV) 
effects in various nuclear processes reveal an amount on
the order of $0.3\%$~\cite{miller90}. This 
is consistent with the expectation that 
CSV effects are caused by electromagnetic interaction and by the
small mass difference between the $u$ and $d$ quarks~\cite{henley1}.

It has been generally assumed that the parton distributions in hadrons
obey charge symmetry. This assumption enables one to relate the parton
distributions in the proton and neutron in a simple fashion,
$u_p(x) = d_n(x), d_p(x) = u_n(x)$, etc. Indeed, charge symmetry is usually 
assumed in the analysis of DIS and DY experiments, which often 
use nuclear targets containing both protons and neutrons. Charge
symmetry is also implicit in the derivation of many QCD sum rules,
including the Gottfried sum rule, the Adler sum rule, and the Bjorken sum rule.

The possibility that charge symmetry could be significantly violated at 
the parton level has been discussed recently by several 
authors~\cite{ma1,ma2,sather,forte93,londergan1,rodionov,benesh1,benesh2}. 
Ma and collaborators~\cite{ma1,ma2} pointed out that 
the violation of the GSR 
can be caused by CSV as well as by flavor asymmetry of the nucleon sea.
They also showed that DY experiments, such as NA51 and E866, are subject
to both flavor asymmetry and CSV effects. 
Forte estimated~\cite{forte93} the relative size of the CSV and flavor
asymmetry via a combined analysis of the GSR violation and the nucleon
$\sigma$ term. Using the bag
model, Rodionov et al~\cite{rodionov} showed  
that a significant CSV effect of $\sim 5$\% could exist 
for the ``minority valence quarks" [i.e. $d_p(x)$ and $u_n(x)$] at large
$x$ ($x > 0.4$). A model study~\cite{benesh2} 
of CSV for sea quarks shows that the effect is
very small, roughly a factor of 10 less than for valence quarks.
Londergan \& Thomas have reviewed the role of CSV for parton 
distributions~\cite{londergan2}.

Recently Boros, Londergan and Thomas (BLT)~\cite{blt98} compared the
parton distributions, $F^{\nu N}_2(x,Q^2)$ and $F^{\mu N}_2(x,Q^2)$. 
The $F^{\mu N}_2 (x,Q^2)$ structure
function extracted from DIS muon scattering is defined as 
\begin{eqnarray}
F^{\mu N}_2(x,Q^2) \equiv {F^{\mu p}_2(x,Q^2) + F^{\mu n}_2 (x,Q^2)
\over 2}
\label{eq:4.5.1}
\end{eqnarray}
\begin{eqnarray}
= {5 \over 18} x [u + \bar u + d + \bar d + {2 \over 5} (s + \bar s)
+ {8 \over 5} (c + \bar c)].
\label{eq:4.5.2}
\end{eqnarray}                                             
In going from Eq.~\ref{eq:4.5.1} to~\ref{eq:4.5.2}, the $(x,Q^2)$
dependence has been suppressed and
charge symmetry ($u_p=d_n=u$, etc.) has been employed. For neutrino DIS one has 
\begin{eqnarray}
F^{\nu N}_2(x,Q^2) = x[u + \bar u + d + \bar d + 2s + 2 \bar c],
\label{eq:4.5.3}
\end{eqnarray}                                             
again using charge symmetry and suppressing the $(x,Q^2)$ 
dependence in the parton
distributions. Neglecting the contribution of  charmed quarks in the
nucleon and correcting for the small difference due to strange quark
contributions one expects
\begin{eqnarray}
{{18 \over 5} F^{\mu N}_2(x,Q^2) \over F^{\nu N}_2(x,Q^2)} \approx 1.
\label{eq:4.5.4}
\end{eqnarray}                                             

After making the necessary small
corrections, BLT found that the ratio in Eq.~\ref{eq:4.5.4} at common 
values of $(x,Q^2)$ is 1 for
$x>0.1$ but for $x< 0.1$ the ratio 
progressively decreases below 1 as $x$
decreases. They suggested that this effect might be due to a violation of
charge symmetry in the sea of the nucleon. 
To achieve agreement with experiment, BLT found it
necessary to set $\bar d_n(x) \approx 1.25 \bar u_p(x)$
and $\bar u_n(x) \approx 0.75 \bar d_p(x)$ at small $x$. 
As the total number of sea quarks is kept 
equal in the neutron and proton ($\bar u_p(x) + \bar d_p(x) \approx u_n(x) +
\bar d_n(x)$), one deduces $\bar d_p(x) \approx \bar u_p(x)$ 
and $\bar d_n(x) \approx (5/3) \bar u_n(x)$, a most striking asymmetry
between the proton and the neutron sea.

How would this claim of large CSV affect the 
E866 analysis of the flavor asymmetry? First, CSV alone could
not account for the E866 data. In fact, an even larger
amount of flavor asymmetry is required to compensate for the possible CSV 
effect~\cite{blt98}.
Second, there has been no indication of CSV for $x > 0.1$. Thus, the
large $\bar d / \bar u$ asymmetry from E866 for $x > 0.1$
is not affected. 
Fortunately, this radical alteration of the
conventional view does not appear to be the case, as shown recently 
by Bodek et al.~\cite{bodek99}, who showed that the measured 
asymmetry of $W^+(W^-)$
production at CDF is consistent with charge symmetry and in strong disagreement
with the suggestion of BLT. The reasons for the discrepancy of the DIS data
with Eq.~\ref{eq:4.5.4} must be found elsewhere.\\

\noindent{\bf{5. FURTHER IMPLICATIONS OF THE MESON-CLOUD MODELS}}

\noindent{\bf{5.1 Strange Sea of the Nucleon}}

Models in which virtual mesons are admitted as degrees of freedom
have implications that extend beyond the $\bar d, \bar u$ 
flavor asymmetry addressed above.
They create hidden strangeness in the nucleon via such virtual processes as 
$p \to \Lambda + K^+, \Sigma + K$, etc. 
Such processes are of considerable interest as they imply different $s$ and
$\bar s$ parton distributions in the nucleon, a feature not found in gluonic
production of $s \bar s$ pairs. This subject has 
received a fair amount of attention
in the literature~\cite{holt96,signal87,warr92,ji95,ma96} 
but experiments have yet to clearly identify such a
difference. Thus in contrast to the $\bar d, \bar u$
flavor asymmetry, to date there is no positive experimental evidence
for $s \bar s$ contributions to the nucleon from virtual 
meson-baryon states~\cite{conrad98,ccfr95}.

A difference between the $s$ and $\bar s$ distribution can be made manifest
by direct measurement of the $s$ and $\bar s$ parton distribution functions 
in DIS neutrino scattering, or in
the measurement of the $q^2$ dependence of the 
strange quark contribution ($F^p_{1s}(q^2)$)
to the proton charge form factor. 
This latter case is not well known and follows from a suggestion of Kaplan 
and Manohar~\cite{kaplan88} regarding the new information contained in the
weak neutral current form factors of the nucleon. Measurement of these form
factors allows extraction of the strangeness contribution to the 
nucleon's charge and magnetic moment and axial form factors. The portion 
of the charge form factor $F^p_{1s} (q^2)$ due to strangeness clearly is
zero at $q^2 = 0$, but if the $s$ and $\bar s$ distributions are different the
form factor becomes non-zero at finite $q^2$. These ``strange'' form 
factors can be measured in neutrino elastic scattering~\cite{garvey93}
from the nucleon, or by selecting the parity-violating component of
electron-nucleon elastic scattering, as is now
being done at the Bates~\cite{mueller} and Jefferson Laboratories~\cite{aniol}.

It is worth pointing out that there is a relationship between the parton
distributions and the form factors of a hadron.  If the neutron's charge 
form factor is explained in terms of a particular meson-baryon expansion,
then one should expect that the expansion is consistant with the neutron''s
partonic structure. Little work appears to have been done bringing these
descriptions together. Below is an example showing the impact of the
nucleon's pionic content, inferred from the flavor asymmetry, on the spin
distribution in the nucleon. \\

\noindent{\bf{5.2 Spin Dependent Structure Functions}}

\begin{figure}
\begin{center}
\psfig{figure=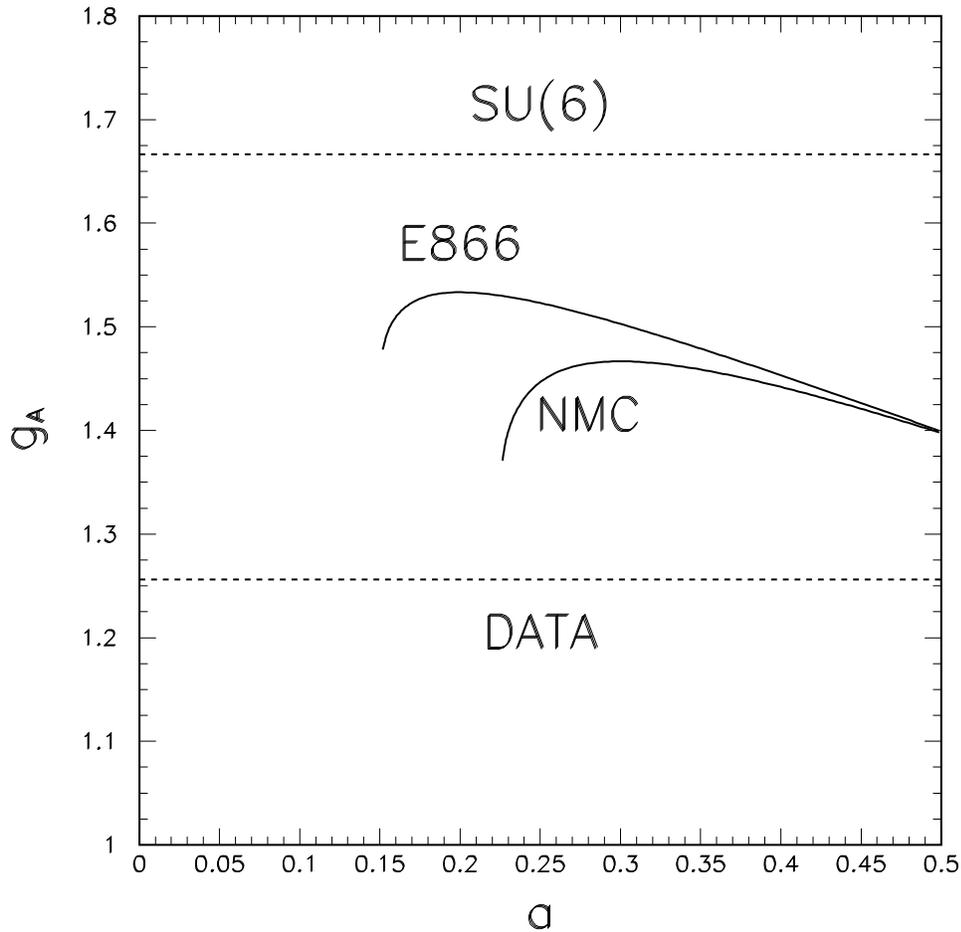,height=5.5in}
\caption{$g_A$ as a function of $a$ determined from Eqs.~\ref{eq:4.2.2}
and~\ref{eq:5.2.1}. The upper and lower solid curve is obtained by using the values
$\bar d - \bar u$ determined by E866 and NMC, respectively. The values
of $g_A$ for the SU(6) limit and from the experiments are also indicated by
the dashed curves.}
\label{fig:5.2.1}
\end{center}
\end{figure}

As the pion is a $J^\pi = 0^-$ hadron it must be emitted 
as a p-wave requiring that the nucleon flip its spin 1/3 of the time
upon emitting a pion. Thus, it appears that such a
process might readily account for the reduction of $g_A$ from 1.667  and the
overall reduction in the quark contribution to nucleon spin as
observed in DIS. The contribution to $g_A$ from
virtual pions is
\begin{eqnarray}
g_A = \Delta u - \Delta d = {5 \over 3} - {20 \over 27}(2a+b) + {32 \over
27} \sqrt{2ab}\\
= 1.53 ~~ {\rm for}~~ a=0.2, b=0.1,
\label{eq:5.2.1}
\end{eqnarray}                                             
and to the total nucleon spin due to up and down quarks
\begin{eqnarray}
\Delta q = \Delta u + \Delta d = 1 - {2 \over 3} (2a -b)\\
= 0.80 ~~{\rm for}~~ a=0.2, b=0.1.
\label{eq:5.2.2}
\end{eqnarray}                                             
Thus, virtual pions do reduce the fraction of nucleon spin carried by up and
down quarks but the effect is not sufficient to reduce 
$g_A$ from its SU(6) value of 5/3 to the experimental value of 1.256. 
Eqs.~\ref{eq:4.2.2} and~\ref{eq:5.2.1}, together with the values
of $\bar d - \bar u$ integral determined from experiments, allow
us to calculate $g_A$ as a function of $a$. The different values of the
$\bar d - \bar u$ integral from the NMC and E866 correspond to two different
curves shown in Figure~\ref{fig:5.2.1}. 
Figure~\ref{fig:5.2.1} illustrates the possible range of $g_A$ 
achievable from virtual pions alone. Presumably relativistic effects are 
responsible for the bulk of that reduction~\cite{thomas88}.
Similarly, virtual pions reduce the total spin carried by
$u$ and $d$ quarks but not enough to be a dominant effect as $\Delta u +
\Delta d \approx 0.25$.

There is an aspect of spin distributions predicted by virtual
pion emission that is present in the data. The spin carried by antiquarks 
in the virtual pion model must be zero. Indeed the measured value~\cite{smc98},
$\Delta \bar q = 0.01 \pm 0.04 \pm 0.03$ is consistent with that picture.
It is very likely that meson will prove to play an important part in
understanding the nucleon's spin structure but that role is not clear 
at present.\\

\noindent{\bf{5.3 Sea Quark Distributions in Hyperons}}

Dilepton production using meson or hyperon beams offers a
means of determining parton distributions of these unstable hadrons. 
Many important features of nucleon parton distributions, such as
the flavor structure and the nature of the non-perturbative sea, find their
counterparts in mesons and hyperons. Information about meson and hyperon 
parton structure could provide valuable new insight into 
nucleon parton distributions. Furthermore,
certain aspects of the nucleon structure, such as the strange 
quark content of the nucleon, could be 
probed with kaon beams.

No data exist for hyperon-induced dilepton production. The observation
of a large $\bar d / \bar u$ asymmetry in the proton has 
motivated Alberg et al.~\cite{alberg1,alberg2}
to consider the sea-quark distributions in the $\Sigma$. The meson-cloud
model implies a $\bar d / \bar u$ asymmetry in the $\Sigma^+$ even larger than
that of the proton. However, the opposite effect is expected
from SU(3) symmetry.
Although relatively intense
$\Sigma^+$ beams have been produced for recent experiments at Fermilab,
this experiment appears to be very challenging because of
large pion, kaon, and proton contaminations in the beam.\\

\noindent{\bf{6. FUTURE PROSPECTS}}

\noindent{\bf{6.1 $\bar d / \bar u$ at Large and Small $x$}}

The interplay between the perturbative and non-perturbative components of
the nucleon sea remains to be better determined. Since the perturbative
process gives a symmetric $\bar d/ \bar u$ while a non-perturbative 
process is needed to generate an asymmetric $\bar d/ \bar u$ sea, the relative
importance of these two components is directly reflected in the $\bar d/ \bar u$
ratios. Thus, it would be very important to extend the DY
measurements to kinematic regimes beyond the current limits. 

\begin{figure}
\begin{center}
\psfig{figure=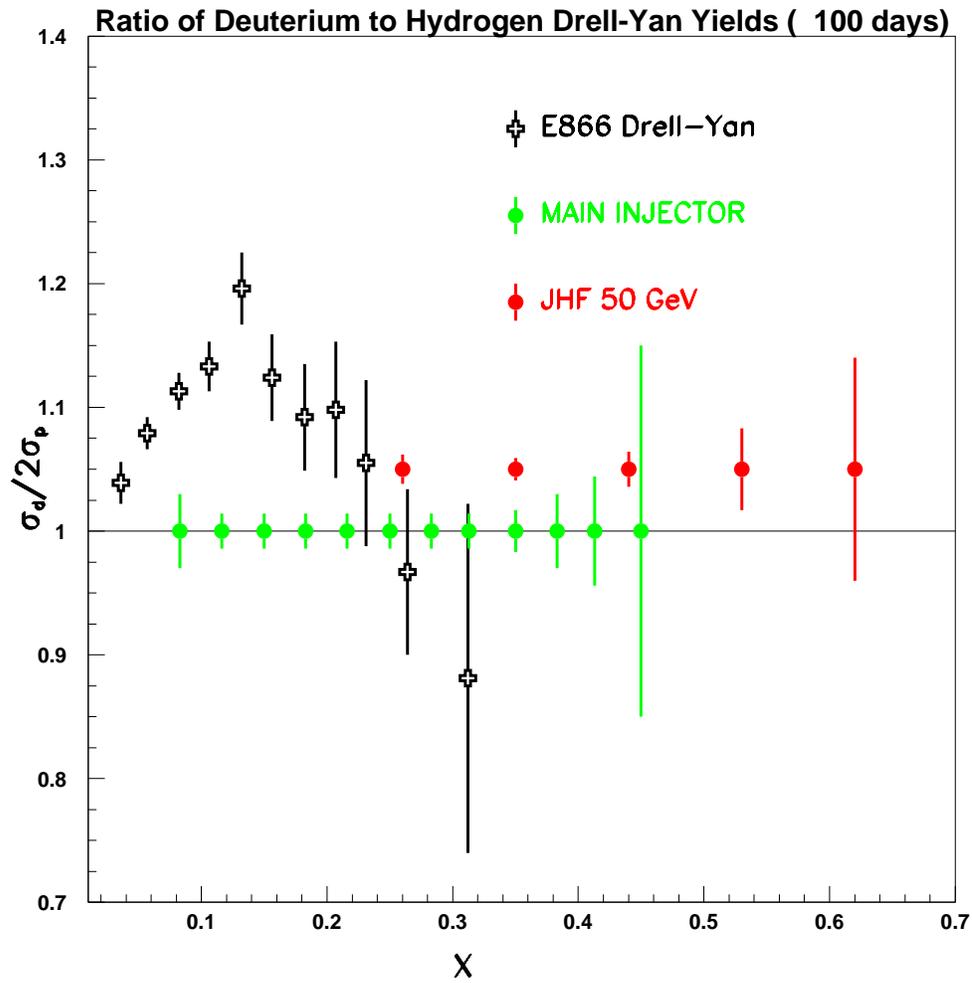,height=5.5in}
\caption{Projected statistical accuracy for $\sigma_d/2\sigma_p$ in a
100-day run at JHF. The E866 data and the projected sensitivity for
a proposed measurement~\cite{p906} at the 120 GeV 
Fermilab Main-Injector are also shown.}
\label{fig:6.1.1}
\end{center}
\end{figure}

The new 120 GeV Fermilab Main Injector (FMI) and the proposed 50 GeV 
Japanese Hadron Facility~\cite{nagamiya} (JHF) present opportunities for 
extending the $\bar d/ \bar u$ measurement to larger $x$ ($x > 0.25$).
For given values of $x_1$ and $x_2$ the DY cross section
is proportional to $1/s$, hence the DY cross section 
at 50 GeV is roughly 16 times greater than
that at 800 GeV! Figure~\ref{fig:6.1.1} shows the expected statistical 
accuracy for $\sigma (p+d)/ 2 \sigma (p+p)$ at JHF~\cite{brown} compared 
with the data from E866 and a proposed measurement~\cite{p906} using 
the 120 GeV proton beam at the FMI. A definitive measurement of
the $\bar d/ \bar u$ over the region $0.25 < x < 0.7$ could indeed be
obtained at FMI and JHF.

At the other end of the energy scale, RHIC will operate soon in the range
$50 \le \sqrt s \le 500$ GeV/nucleon. The capability of accelerating and
colliding a variety of beams from $p + p$, $p + A$, to $A + A$ at RHIC
offers a unique opportunity to extend the DY $\bar d / \bar u$ measurement
to very small $x$.\\

\noindent{\bf{6.2 W Production}}

\begin{figure}
\begin{center}
\psfig{figure=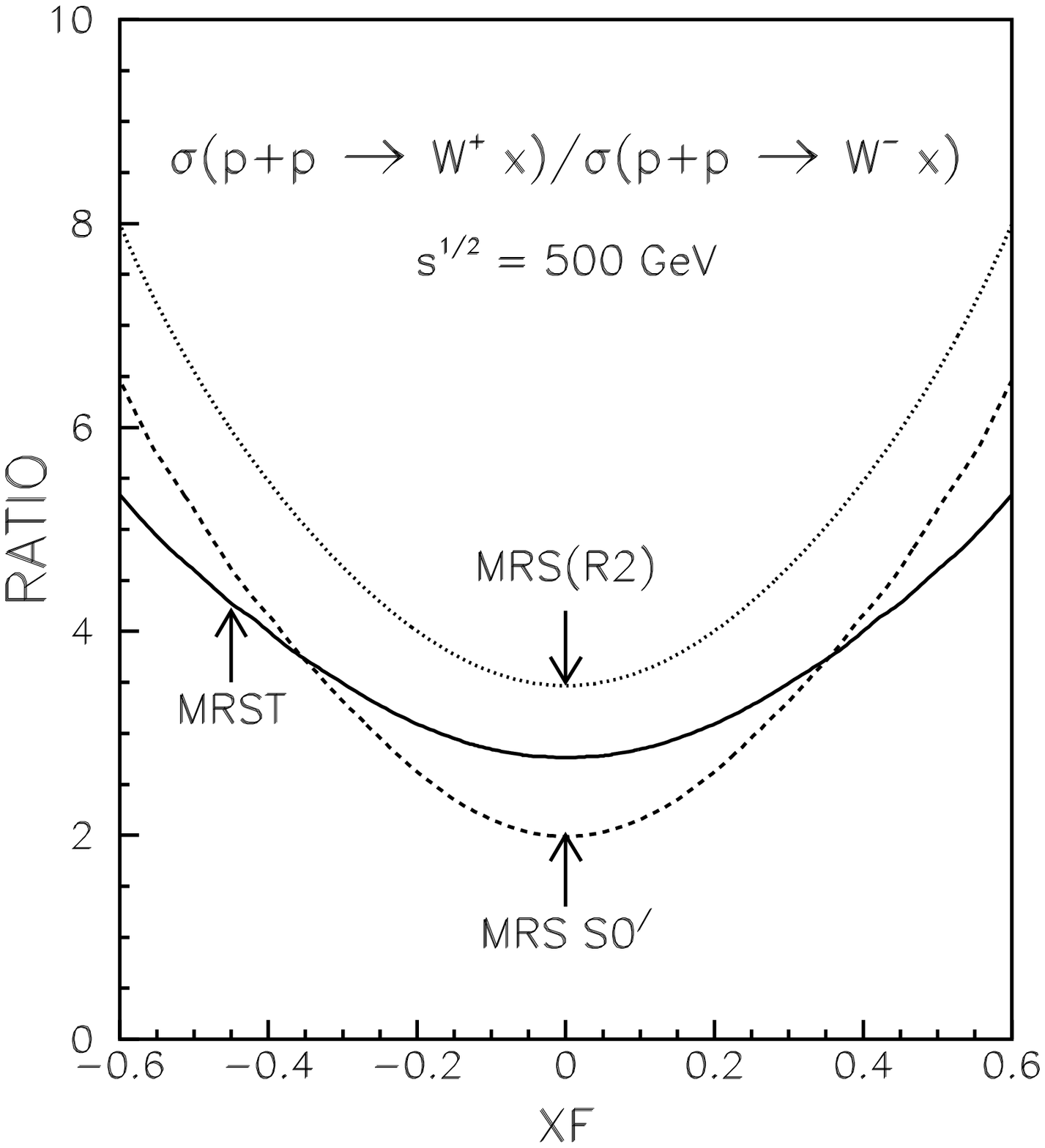,height=5.5in}
  \caption{ Predictions of 
  $\sigma (p+p \to W^+ X) / \sigma (p+p \to W^- X)$ as a function of $x_F$
  at $\sqrt s$ = 500 GeV.
  The dashed curve corresponds to the $\bar
  d/\bar u$ symmetric MRS S0$^\prime$ structure 
  functions, while the solid and dotted curves
  are for the $\bar d/\bar u$ asymmetric structure function MRST and MRS(R2),
  respectively.}
\label{fig:6.2.1}
\end{center}
\end{figure}

To disentangle the $\bar d / \bar u$ asymmetry from the possible 
CSV effect, one could consider $W$ boson production, a generalized
DY process, in $p + p$ collision at RHIC.
An interesting quantity to be measured is the ratio of the 
$p + p \to W^+ + X$ and $p + p \to W^- + X$ cross sections~\cite{peng1}. 
It can be shown that this
ratio is very sensitive to $\bar d / \bar u$. An important feature of
the $W$ production asymmetry in $p + p$ collision is that it is completely free 
from the assumption of charge symmetry. Figure~\ref{fig:6.2.1} shows the 
predictions for $p + p$ collision at $\sqrt s =
500~$GeV. The dashed curve corresponds to the $\bar
d/\bar u$ symmetric MRS S0$^\prime$~\cite{mrss0} structure 
functions, while the solid and dotted curves
are for the $\bar d/\bar u$ asymmetric structure function MRST and MRS(R2),
respectively. Figure~\ref{fig:6.2.1} clearly shows that $W$ asymmetry 
measurements at RHIC could provide an independent determination 
of $\bar d / \bar u$.\\

\noindent{\bf{6.3 Strange Sea in the Nucleon}}

As discussed earlier, an interesting consequence of the meson-cloud model
is that the $s$ and $\bar s$ distributions
in the proton could have very different shapes, even though the net amount
of strangeness in the proton vanishes. By 
comparing the $\nu$ and $\bar \nu$
induced charm production, the CCFR 
collaboration found no difference between 
the $s$ and $\bar s$ distributions~\cite{ccfr95}. More precise
future measurements would be very helpful.
Dimuon production experiments using $K^\pm$ beams might provide an independent
determination of the $s$/$\bar s$ ratio of the proton, provided that our
current knowledge on valence-quark distributions in kaons is improved.
As discussed in Section 5.1, ongoing measurements of $F^p_{1s}$ via 
parity-violating electron-nucleon scattering should shed much light on
the possible difference between $s$ and $\bar s$ distributions.

\noindent{\bf{6.4 Sea Quark Polarization}}

Polarized DY and $W^\pm$ production in polarized $p+p$ collision 
are planned at RHIC~\cite{bunce} and they have great potential for providing
qualitatively new information about antiquark polarization. At large
$x_F$ region ($x_F > 0.2$), the longitudinal spin asymmetry $A_{LL}$ in the
$p+p$ DY process is given by~\cite{moss,plm}
\begin{eqnarray}
A^{DY}_{LL}(x_1,x_2) \approx g_1(x_1)/F_1(x_1) \times {\Delta \bar u
\over \bar u}(x_2),
\label{eq:6.4.1}
\end{eqnarray}  
where $g_1(x)$ is the proton polarized structure function measured in DIS,
and $\Delta \bar u(x)$ is the polarized $\bar u$ distribution function.

Eq.~\ref{eq:6.4.1} shows that $\bar u$ polarization can be determined using 
polarized DY at RHIC. Additional information on the sea-quark polarization
can be obtained via $W^\pm$ production~\cite{bs93}.
The parity-violating
nature of $W$ production implies that only one of the two beams need to
be polarized. At positive $x_F$ (along the direction of the polarized
beam), one finds\cite{bs93},
\begin{eqnarray}
A_L^{W^+}\approx {\Delta u \over u}(x_2),\ \ {\rm and}\ \ \
A_L^{W^-}\approx {\Delta d\over d}(x_2),  \label{eq:6.4.2} 
\end{eqnarray}
where $A^W_L$ is the single-spin asymmetry for $W$ production.
Eq.~\ref{eq:6.4.2} shows that the flavor dependence of the sea-quark
polarization can be revealed via $W^\pm$ production at RHIC.
A remarkable prediction of the chiral quark-soliton model is that
the flavor asymmetry of polarized sea-quark, $\Delta \bar u(x) -
\Delta \bar d(x)$, is very large~\cite{dres99a}. This is in striking
contrast to the meson cloud model which predicts very small 
values for $\Delta \bar u(x) - \Delta \bar d(x)$~\cite{fries98,boreskov99}.
Future DY and $W^\pm$ production experiments at RHIC could clearly test
these models~\cite{dres99b}\\

\noindent{\bf{7. CONCLUSION}}

The flavor asymmetry of the nucleon sea has been clearly established
by recent DIS and DY experiments. The surprisingly large asymmetry
between $\bar u$ and $\bar d$ is unexplained by perturbative
QCD. Thus far, three distinct explanations have been offered in the 
literature for the origin of the observed $\bar d$, $\bar u$ asymmetry. 
The first is Pauli blocking which, while appealing, is difficult to calculate 
and appears to produce too small an effect to be the sole origin of 
the large observed asymmetry. 
The second involves virtual isovector mesons, mostly pions, in the nucleon. 
Such descriptions necessarily require non-perturbative QCD and, apart from 
lattice gauge calculations, demand additional parameters and possess 
systematic uncertainties. However, as these virtual mesons readily generate 
a large $\bar d$, $\bar u$ asymmetry, many authors, using a variety of 
techniques to invoke and justify their approachs, have obtained qualitative 
agreement with the experimental measurements. The third explanation involves 
instantons, but the theory is not sufficiently developed to allow 
quantitative comparison to the asymmetry data.
Future experiments will test and refine these models. They will further
illuminate the interplay between the perturbative and non-perturbative
nature of the nucleon sea.

\vskip 0.4in
\noindent{Acknowledgment}

We are grateful to our collaborators on the Fermilab E772 and E866.
This work 
was supported by the US Department of Energy, Nuclear Science 
Division, under contract W-7405-ENG-36.

\newpage

\begin{table}[tbh]
\caption {Values of the integral $\int_0^1 (\bar d(x) - \bar u(x)) dx$ 
for various parton distribution functions.}
\begin{center}
\begin{tabular}{|c|c|}
\hline
PDF & integral \\
\hline\hline
CTEQ4M & 0.108 \\
\hline
MRS(R2) & 0.162 \\
\hline
GRV94 & 0.163 \\
\hline
CTEQ5M & 0.124 \\
\hline
MRST & 0.102 \\
\hline
GRV98 & 0.126 \\
\hline
\end{tabular}
\end{center}
\label{tab:3.6.1} 
\end{table}
\newpage

\begin{table}[tbh]
\caption {Comparison between the 450 GeV/c NA51 result and the 800 GeV/c 
E866 data point near $x_2 = 0.18$.}
\begin{center}
\begin{tabular}{|c|c|c|c|c|c|}
\hline
& $\langle x_2\rangle$ & $\langle x_F\rangle$ & 
$\langle M_{\mu \mu}\rangle$(GeV) & $\sigma^{pd}/ 2 \sigma^{pp}$ & 
$\bar d/ \bar u$ \\
\hline\hline
NA51 & 0.18 & 0.0 & 5.2 & 1.099 $\pm$ 0.039 & 1.96 $\pm$ 0.246 \\
\hline
E866 & 0.182 & 0.192 & 9.9 & 1.091 $\pm$ 0.044 & 1.41 $\pm$ 0.146 \\
\hline
\end{tabular}
\end{center}
\label{tab:3.7.1} 
\end{table}

\newpage
\vfill
\eject

\vfill
\eject

\vfill
\eject

\vfill
\eject

\vfill
\eject

\vfill
\eject

\vfill
\eject

\vfill
\eject

\vfill
\eject

\vfill
\eject

\vfill
\eject

\vfill
\eject

\vfill
\eject

\vfill
\eject

\vfill
\eject

\newpage

\begin{thebibliography}{99}

\bibitem {bloom69} E. D. Bloom et al., Phys. Rev. Lett. {\bf 23} (1969) 930.

\bibitem {breiden69} M. Breidenbach et al., Phys. Rev. Lett. {\bf 23} (1969)
935.

\bibitem {feyn72} R. P. Feynman, Photon-Hadron Interactions (Benjamin,
New York, 1972).

\bibitem {liu99} K. F. Liu et al., Phys. Rev. {\bf D59} (1999) 112001.

\bibitem {bj69} J. D. Bjorken, Phys. Rev. {\bf 179} (1969) 1547.

\bibitem {feyn69} R. P. Feynman, Phys. Rev. Lett. {\bf 23} (1969) 1415.

\bibitem {drell69} S. D. Drell, D. J. Levy and T. M. Yan,
Phys. Rev. {\bf 187} (1969) 2159.

\bibitem {drell70} S. D. Drell, D. J. Levy and T. M. Yan,
Phys. Rev. {\bf D1} (1970) 1035; {\bf D1} (1970) 1617.

\bibitem {cab70} N. G. Cabbibo, G. Parisi, M. Testa and A. Verganelakis,
Lett. Nuovo Cimento {\bf 4} (1970) 569.

\bibitem {lee72} T. D. Lee and S. D. Drell, Phys. Rev. {\bf D5} (1972) 1738.

\bibitem {bj69a} J. D. Bjorken and E. A. Paschos, Phys. Rev. {\bf 185}
(1969) 1975.

\bibitem {kuti71} J. Kuti and V. F. Weisskopf,
Phys. Rev. {\bf D4} (1971) 3418.

\bibitem {land71} P. V. Landschoff and J. C. Polkinghorne,
Nucl. Phys. {\bf B28} (1971) 240.

\bibitem {gell64} M. Gell-Mann, Phys. Lett. {\bf 8} (1964) 214.

\bibitem {zweig64} G. Zweig, CERN Preprint 8182/TH401, 8419/TH412 (1964).

\bibitem {friedman} J. I. Friedman and H. W. Kendall, Annu. Rev.
Nucl. Sci. {\bf 22} (1972) 203.

\bibitem {cdhs79} J. G. H. de Groot et al., Z. Phys. {\bf C1} (1979) 143.

\bibitem {abrom82} H. Abromowicz et al., Z. Phys. {\bf C15} (1982) 19.

\bibitem {conrad98} J. M. Conrad, M. H. Shaevitz and T. Bolton, Rev. Mod. Phys.
{\bf 70} (1998) 1341.

\bibitem {gott67} K. Gottfried, Phys. Rev. Lett. {\bf 18} (1967) 1174.

\bibitem {bj67} J. D. Bjorken, Proc. 1967 Int. Sym. on
Electron and Photon Interactions at High Energies, Stanford, CA (1967) 109.

\bibitem {bloom70} E. D. Bloom et al., Proc. 15th Int. Conf.
on High Energy Phys., Kiev, USSR (1970).

\bibitem {bloom73} E. D. Bloom, Proc. 6th Int. Sym. on 
Electron and Photon Interactions at High Energies, 
edited by H. Rollnik and W. Pfeil (North-Hollan, Amsterdam, 1974) 227.

\bibitem {field77} R. D. Field and R. P. Feynman,
Phys. Rev. {\bf D15} (1977) 2590.

\bibitem {wate75} Y. Watenabe et al., Phys. Rev. Lett. {\bf 35} (1975) 898.

\bibitem {chang75} C. Chang et al., Phys. Rev. Lett. {\bf 35} (1975) 901.

\bibitem {chio79} B. A. Gordon et al., Phys. Rev. {\bf D20} (1979) 2645.

\bibitem {emc87} J. J. Aubert et al., Nucl. Phys. {\bf B293} (1987) 740.

\bibitem {bcdms90} A. C. Benvenuti et al., Phys. Lett. {\bf B237} (1990) 599.

\bibitem {bodek79} A. Bodek et al., Phys. Rev. {\bf D20} (1979) 1471.

\bibitem {do84} D. W. Duke and J. F. Owens, Phys. Rev. {\bf D30} (1984) 49.

\bibitem {ehlq84} E. Eichten, I. Hinchliffe, K. Lane and C. Quigg,
Rev. Mod. Phys. {\bf 56} (1984) 579;
Rev. Mod. Phys. {\bf 58} (1985) 1065.

\bibitem {dflm88} M. Diemoz, F. Ferroni, E. Longo and G. Martinelli,
Z. Phys. {\bf C39} (1988) 21.

\bibitem {mrs88} A. D. Martin, R. G. Roberts and W. J. Stirling,
Phys. Lett. {\bf B206} (1988) 327.

\bibitem {abfow89} P. Aurenche et al., Phys. Rev. {\bf D39} (1989) 3275.

\bibitem {chris70} J. H. Christensen et al.,
Phys. Rev. Lett. {\bf 25} (1970) 1523.

\bibitem {dy71} S. D. Drell and T. M. Yan, Phys. Rev. Lett. {\bf 25} (1970) 316.

\bibitem {herb77} S. W. Herb et al., Phys. Rev. Lett. {\bf 39} (1977) 252.

\bibitem {kaplan77} D. M. Kaplan et al., Phys. Rev. Lett. {\bf 40} (1977) 435.

\bibitem {ito81} A. S. Ito et al., Phys. Rev. {\bf D23} (1981) 604.

\bibitem {smith81} S. R. Smith et al., Phys. Rev. Lett. {\bf 46} (1981) 1607.

\bibitem {emc83} J. J. Aubert et al., Phys. Lett. {\bf B123} (1983) 295.

\bibitem {emc88} J. Ashman et al., Phys. Lett. {\bf B202} (1988) 603.

\bibitem {nmc90} D. Allasia et al., Phys. Lett. {\bf B249} (1990) 366.

\bibitem {nmc95} P. Amaudruz et al., Nucl. Phys. {\bf B441} (1995) 3.

\bibitem {nmc91} P. Amaudruz et al., Phys. Rev. Lett. {\bf 66} (1991) 2712.

\bibitem {slac92} L. W. Whitlow et al., Phys. Lett. {\bf B282} (1992) 475.

\bibitem {bcdms90a} A. C. Benvenuti et al., Phys. Lett. {\bf B237} (1990) 592.

\bibitem {emc90} M. Arneodo et al., Nucl. Phys. {\bf B333} (1990) 1.

\bibitem {nmc94} M. Arneodo et al., Phys. Rev. {\bf D50} (1994) R1.

\bibitem {nmc92} P. Amaudruz et al., Phys. Lett. {\bf B295} (1992) 159.

\bibitem {nmc97} M. Arneodo et al., Nucl. Phys. {\bf B487} (1997) 3.

\bibitem {nmc95a} M. Arneodo et el., Phys. Lett. {\bf B364} (1995) 107.

\bibitem {nmc97a} M. Arneodo et al., Nucl. Phys. {\bf B483} (1997) 3.

\bibitem {hinch96} I. Hinchliffe and A. Kwiatkowski,
Annu. Rev. Nucl. Part. Sci. {\bf 46} (1996) 609.

\bibitem {ross79} D. A. Ross and C. T. Sachrajda, Nucl. Phys.
{\bf B149} (1979) 497.

\bibitem {kotaev96} A. L. Kotaev, A. V. Kotikov, G. Parente,
and A. V. Sidorov, Phys. Lett.
{\bf B388} (1996) 179.

\bibitem {forte94a} R. D. Ball and S. Forte, Nucl. Phys.
{\bf B425} (1994) 516.

\bibitem {forte94b} R. D. Ball, V. Barone, S. Forte, and
M. Genovese, Phys. Lett.
{\bf B329} (1994) 505.

\bibitem {gee95} D. F. Geesaman, K. Saito and A. W. Thomas,
Annu. Rev. Nucl. Part. Sci. {\bf 45} (1995) 337.

\bibitem {pionex83} C. H. Llewllyn-Smith, Phys. Lett.
{\bf B128} (1983) 107; M. Ericson and A. W. Thomas, Phys. Lett.
{\bf B128} (1983) 112; E. L. Berger, F. Coester and R. B. Wiringa,
Phys. Rev. {\bf D29} (1984) 398.

\bibitem {e772a} D. M. Alde et al., Phys. Rev. Lett.
{\bf 64} (1990) 2479.

\bibitem {close85} F. E. Close, R. L. Jaffe, R. G. Roberts and
G. G. Ross, Phys. Rev. {\bf D31} (1985) 1004.

\bibitem {e772b} P. L. McGaughey et al., Phys. Rev. Lett.
{\bf 69} (1991) 1726.

\bibitem {es91} S. D. Ellis and W. J. Stirling, Phys. Lett.
{\bf B256} (1991) 258.

\bibitem {ehq92} E. Eichten, I. Hinchliffe and C. Quigg,
Phys. Rev. {\bf D45} (1992) 2269.

\bibitem {mrst} A. D. Martin, R. G. Roberts, W. J. Stirling and
R. S. Thorne, Eur. Phys. J. {\bf C4} (1998) 463.

\bibitem {kumano95} S. Kumano, Phys. Lett. {\bf B342} (1995) 339.

\bibitem {na51} A. Baldit et al., Phys. Lett. {\bf B332} (1994) 244.

\bibitem{e866} E. A. Hawker et al., Phys. Rev. Lett. {\bf 80} (1998) 3715.

\bibitem{towell} R. S. Towell et al., to be published (1999).

\bibitem{cteq} H. L. Lai et al., Phys. Rev. {\bf D55} (1997) 1280.

\bibitem{mrs} A. D. Martin, R. G. Roberts and W. J. Stirling, Phys.
Lett. {\bf B387} (1996) 419.

\bibitem{peng98} J. C. Peng et al., Phys. Rev. {\bf D58} (1998) 092004.

\bibitem {grz73} M. Gronau, F. Ravndal and Y. Zarmi, Nucl. Phys.
{\bf B51} (1973) 611.

\bibitem {ashman91} J. Ashman et al., Z. Phys. {\bf C52} (1991) 361.

\bibitem {bebc84} D. Allasia et al., Phys. Lett. {\bf B135} (1984) 231.

\bibitem {cdhs84} H. Abramowicz et al., Z. Phys. {\bf C25} (1984) 29.

\bibitem {levelt} J. Levelt, P. J. Mulders and A. W. Schreiber,
Phys. Lett. {\bf B263} (1991) 498.

\bibitem {hermes98} K. Ackerstaff et al., Phys. Rev. Lett.
{\bf 81} (1998) 5519.

\bibitem {grv94} M. Gl\"{u}ck, E. Reya and A. Vogt, Z. Phys.
{\bf C67} (1995) 433.

\bibitem {frank89} L. L. Frankfurt et al., Phys. Lett.
{\bf B230} (1989) 141.

\bibitem {close91} F. E. Close and R. G. Milner,
Phys. Rev. {\bf D44} (1991) 3691.

\bibitem {smc98} B. Adeva et al., Phys. Lett. {\bf B369} (1997) 93;
B. Adeva et al., Phys. Lett. {\bf 420} (1998) 180.

\bibitem {hermes99} K. Akerstaff et al., hep-ex/9906035 (1999).

\bibitem {grv98} M. Gl\"{u}ck, E. Reya and A. Vogt, Eur. Phys. J.
{\bf C5} (1998) 461.

\bibitem {cteq5} H. L. Lai et al., hep-ph/9903282, to be published
in Eur. Phys. J. (1999).

\bibitem {lai99a} H. L. Lai, private communication (1999).

\bibitem {cdf98} F. Abe et al., Phys. Rev. Lett.
{\bf 81} (1998) 5754.

\bibitem {ball94} R. D. Ball et al., Phys. Lett. {\bf B329} (1994) 505.

\bibitem {stef97} F. M. Steffens and A. W. Thomas, Phys. Rev.
{\bf C55} (1997) 900.

\bibitem {thomas83} A. W. Thomas, Phys. Lett.
{\bf B126} (1983) 97.

\bibitem {sullivan} J. D. Sullivan, Phys. Rev. {\bf D5} (1972) 1732.

\bibitem {kumano98} S. Kumano, Phys. Rep. {\bf 303} (1998) 183.

\bibitem {speth98} J. P. Speth and A. W. Thomas, Adv. Nucl. Phys.
{\bf 24} (1998) 83.

\bibitem{henley} E. M. Henley and G. A. Miller, Phys. Lett. {\bf 251}
(1990) 453. 

\bibitem{kumano} S. Kumano, Phys. Rev. {\bf D43} (1991) 3067; {\bf D43}
(1991) 59; S. Kumano and J. T. Londergan, Phys. Rev. {\bf D44} (1991) 717.

\bibitem{signal} A. Signal, A. W. Schreiber, and A. W. Thomas,
Mod. Phys. Lett. {\bf A6} (1991) 271.

\bibitem{hwang} W. P. Hwang, J. Speth and G. E. Brown, Z. Phys. {\bf A339}
(1991) 383.

\bibitem{szczurek} A. Szczurek, J. Speth and G. T. Garvey, Nucl. Phys.
{\bf A570} (1994) 765.

\bibitem{koepf} W. Koepf, L. L. Frankfurt and M. Strikman,
Phys. Rev. {\bf D53} (1996) 2586.

\bibitem {wally98} W. Melnitchouk, J. Speth and A. W. Thomas,
Phys. Rev. {\bf D59} (1999) 014033.

\bibitem {holt96} H. Holtzman, A. Szczurek and J. Speth,
Nucl. Phys. {\bf A596} (1996) 631.

\bibitem {niko99} N. N. Nikolaev, W. Schaefer, A. Szczurek and
J. Speth, Phys. Rev. {\bf D60} (1999) 014004.

\bibitem {magnin99} J. Magnin and H. R. Christiansen, hep-ph/9903440 (1999).

\bibitem {waka98} M. Wakamatsu and T. Kubota, Phys. Rev. {\bf D57} (1998) 5755.

\bibitem {poby99} P. V. Pobylitsa et al., Phys. Rev. {\bf D59} (1999) 034024.

\bibitem {diak96} D. I. Diakonov et al., Nucl. Phys. {\bf B480} (1996) 341.

\bibitem {diak97} D. I. Diakonov et al., Phys. Rev. {\bf D56} (1997) 4069.

\bibitem {li95} T. P. Cheng and L. F. Li, Phys. Rev. Lett. {\bf 74} (1995) 2872.

\bibitem {li97} T. P. Cheng and L. F. Li, Phys. Lett. {\bf B366} (1996) 365.

\bibitem {weise98} K. Suzuki and W. Weise, Nucl. Phys.
{\bf A634} (1998) 141.

\bibitem {szc96} A. Szczurek, A. Buchmans and A. Faessler,
J. Phys. {\bf C22} (1996) 1741.

\bibitem {song97} X. Song, J. S. McCarthy and H. J. Weber,
Phys. Rev. {\bf D55} (1997) 2624.

\bibitem {ohlsson99} T. Ohlsson and H. Snellman,
Eur. Phys. J. {\bf C7} (1999) 501.

\bibitem {georgi84} A. Manohar and H. Georgi, Nucl. Phys.
{\bf B234} (1984) 189.

\bibitem {bel75} A. A. Belavin et al., Phys. Lett. {\bf B59} (1975) 85.

\bibitem {hooft76} G. t''Hooft,  Phys. Rev. Lett. {\bf 37} (1976) 8.

\bibitem {shu98} T. Schaefer and E. Shuryak, Rev. Mod. Phys. 
{\bf 70} (1998) 323.

\bibitem {forte89} S. Forte, Phys. Lett.
{\bf B224} (1989) 189.

\bibitem {forte91} S. Forte, Acta Phys. Polon.
{\bf B22} (1991) 1065.

\bibitem {inst93} A. E. Dorokhov and N. I. Kochelev, Phys. Lett. {\bf B259}
(1991) 335; {\bf B335} (1993) 167.

\bibitem {miller90} G. A. Miller, B. M. K. Nefkens and I. Slaus,
Phys. Rep. {\bf 194} (1990) 1.

\bibitem{henley1} E. M. Henley and G. A. Miller, {\it Mesons in Nuclei},
ed. M. Rho, D. H. Wilkinson. Amsterdam: North-Holland (1979).

\bibitem{ma1} B-Q. Ma, Phys. Lett. {\bf B274} (1992) 111.

\bibitem{ma2} B-Q. Ma, A. Sch\"{a}fer and  W. Greiner, Phys. Rev. {\bf D47}
(1993) 51.

\bibitem{sather} E. Sather, Phys. Lett. {\bf B274} (1992) 433.

\bibitem {forte93} S. Forte, Phys. Rev.
{\bf D47} (1993) 1842.

\bibitem{londergan1} J. T. Londergan et al., Phys. Lett. {\bf B340} (1994)
115.

\bibitem{rodionov} E. N. Rodionov, A. W. Thomas and J. T. Londergan,
Int. J. Mod. Phys. Lett. {\bf A9} (1994) 1799.

\bibitem{benesh1} C. J. Benesh and T. Goldman, Phys. Rev. {\bf C55} (1997) 441.

\bibitem{benesh2} C. J. Benesh and J. T. Londergan, Phys. Rev. {\bf C58}
(1998) 1218.

\bibitem{londergan2} J. T. Londergan and A. W. Thomas,
{\it Progress in Particle and Nuclear Physics}, {\bf 41} (1998) 49.

\bibitem {blt98} C. Boros, J. T. Londergan and A. W. Thomas, Phys. Rev. Lett. 
{\bf 81} (1998) 4075; Phys. Rev. {\bf D59} (1999) 074021.

\bibitem {bodek99} A. Bodek et al., Phys. Rev. Lett. {\bf 83} (1999) 2892.

\bibitem {signal87} A. I. Signal and A. W. Thomas, Phys. Lett. 
{\bf B191} (1987) 205.

\bibitem {warr92} M. Burkardt and B. J. Warr, Phys. Rev. {\bf D45} (1992)
958.

\bibitem {ji95} X. D. Ji and J. Tang, Phys. Lett. {\bf B362} (1995) 182.

\bibitem {ma96} S. Brodsky and B-Q. Ma, Phys. Lett. {\bf B381} (1996) 317.

\bibitem {ccfr95} A. O. Bazarko et al., Z. Phys. {\bf C65} (1995) 189.

\bibitem {kaplan88} D. Kaplan and A. Manohar, Nucl. Phys. {\bf B310} (1988) 527.

\bibitem {garvey93} G. Garvey, W. Louis and H. White, Phys. Rev. {\bf C48}
(1993) 761.

\bibitem {mueller}  B. Mueller et al., Phys. Rev. Lett. {\bf 78} (1997) 3824.

\bibitem {aniol} K. Aniol et al., Phys. Rev. Lett. {\bf 82} (1999) 1096.

\bibitem {thomas88} A. W. Schreiber and A. W. Thomas, Phys. Lett. {\bf B215}
(1998) 141.

\bibitem{alberg1} M. Alberg et al., Phys. Lett. {\bf B389} (1996) 367.

\bibitem{alberg2} M. Alberg, T. Falter and E. M. Henley,
Nucl. Phys. {\bf A644} (1998) 93.

\bibitem{nagamiya} JHF Project Office. 
{\it Proposal for Japan Hadron Facility},
KEK Rep. {\bf 97-3} (1997).

\bibitem{brown} C. N. Brown, unpublished (1998).

\bibitem{p906} D. Geesaman et al., {\it Fermilab proposal P906} (1999). 

\bibitem{peng1} J. C. Peng and  D. M. Jansen, Phys. Lett. B {\bf 354} 
(1995) 460.

\bibitem{mrss0} A. D. Martin, W. J. Stirling and R. G. Roberts, Phys.
                Lett. B {\bf 306} (1993) 145.

\bibitem{bunce} G. Bunce et al. Part. World {\bf 3} (1992) 1.

\bibitem{moss} J. M. Moss, {\it Int. Conf. 
Polarization Phenomena in Nucl. Phys., AIP Conf. Proc.} 
{\bf 339} (1994) 721.

\bibitem{plm} P. L. McGaughey, J. M. Moss and J. C. Peng, 
Annu. Rev. Nucl. Part. Sci. {\bf 49} (1999) 217.

\bibitem{bs93} C. Bourrely and J. Soffer Phys.  Lett. {\bf B314} (1993)  
132.

\bibitem{dres99a} B. Dressler, K. Goeke, M. V. Polyakov and
C. Weiss, hep-ph/9909541 (1999).

\bibitem{fries98} R. J. Fries and A. Sch\"{a}fer, 
Phys. Lett. {\bf B443} (1998) 40.

\bibitem{boreskov99} K. G. Boreskov and A. B. Kaidalov, Eur. Phys. J.
{\bf C10} (1999) 143.

\bibitem{dres99b} B. Dressler et al., hep-ph/9910464 (1999).

\end{thebibliography}
\end{document}